# CTVR-EHO TDA-IPH Topological Optimized Convolutional Visual Recurrent Network for Brain Tumor Segmentation and Classification

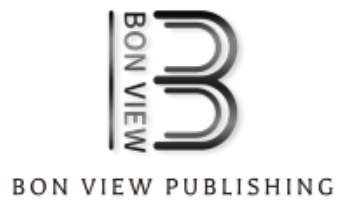

Dhananjay Joshi[1*], Bhupesh Kumar Singh[2], Kapil Kumar Nagwanshi[3] Senior Member, IEEE , Nitin S. Choubey[4]

[1]Research Scholar, Amity School of Engineering and Technology, Amity University Rajasthan (AUR), Jaipur, India.
[2]Amity School of Engineering and Technology, Amity University Rajasthan (AUR), Jaipur, India.
[3]Computing Science & Engineering, Guru Ghasidas Vishwavidyalaya (A Central University) Bilaspur, India
[4]Department of Computer Science, Mukesh Patel School of Technology Management and Engineering Shirpur, SVKM's NMIMS (Deemed to be University), Mumbai, India

*Corresponding author: Dhananjay Joshi, Amity School of Engineering and Technology, Amity University Rajasthan (AUR), Jaipur, India. Email: dj4query@gmail.com

Abstract: In today's world of healthcare, brain tumor detection has become increasingly prevalent. However, the manual brain tumor classification of brain tumors is a time-consuming process. Consequently, Deep Convolutional Neural Network (DCNN) is used by many researchers in the medical field for making accurate diagnoses and aiding in the patient treatment. The traditional techniques have problems such as overfitting and the inability to extract necessary features. To address these issues, we developed the Topological Data Analysis based Improved Persistent Homology (TDA-IPH) and Convolutional Transfer learning and Visual Recurrent learning with Elephant Herding Optimization hyper-parameter tuning (CTVR-EHO) models for brain tumor segmentation and classification. Initially, the Topological Data Analysis based Improved Persistent Homology is designed to segment the brain tumor image. Then, from the segmented image, features are extracted using Transfer learning via the AlexNet model and Bidirectional Visual Long Short-Term Memory (Bi-VLSTM). Next, Elephant Herding Optimization is used to tune the hyperparameters of both networks to get an optimal result. Finally, extracted features are concatenated and classified using the softmax activation layer. The simulation result of this proposed CTVR-EHO and TDA-IPH method is analyzed based on precision, accuracy, recall, loss, and F score metrics. Compared to other existing brain tumor segmentation and classification models, the proposed CTVR-EHO and TDA-IPH approaches show high accuracy (99.8%), high recall (99.23%), high precision (99.67%), and high F score (99.59%).

Keywords: AlexNet, Brain tumor, Deep Convolutional Neural Network (DCNN), Elephant Herding Optimization (EHO), Topological Data Analysis (TDA).

## 1. Introduction

Brain tumors (BT) have been a major cause of mortality worldwide in recent years The occurrence of brain tumors has been rising, with a growing number of diagnoses attributed to advances in medical imaging and awareness. BTs are generally classified into two main types: primary and secondary. Primary brain tumors originate within the brain itself, are often benign, and make up around 70% of brain malignancies, typically staying localized within the brain. In contrast, secondary or metastatic tumors reach the brain through the bloodstream from cancers in other organs, such as the breast, kidney, or lungs [1]. Malignant brain tumors are highly aggressive, affecting individuals of all ages and presenting a considerable risk of mortality. They are estimated to cause 3.6 deaths per 100,000 females and 5.4 deaths per 100,000 males annually from 2014 to 2018. The World Health Organization classified brain tumors into four grades (I to IV) in 2021. Each higher grade of brain tumor has a worse prognosis and is more malignant. Glioblastoma is the most aggressive primary brain tumor, classified as a Grade IV tumor, and has a median survival period of only 12 to 15 months post-diagnosis [2]. Additional kinds of primary brain tumors include gliomas (G), pituitary tumors (P), and meningiomas (M). Gliomas, which arise from glial cells, vary from low-grade tumors (such as oligodendrogliomas) to high-grade tumors (like glioblastomas) and are known for their infiltrative growth patterns. Pituitary tumors, often benign adenomas, can impact hormone production, leading to significant endocrine imbalances and various clinical symptoms, including headaches and vision problems. Meningiomas, typically benign, develop from the brain's protective layers and the spinal cord; however, they can cause complications based on their location and size, leading to symptoms such as seizures or neurological deficits. The variations in tumor types, sizes, and locations makes accurate diagnosis challenging before treatment begins [3]. To aid in diagnosis, advanced imaging techniques like MRI play a critical role in identifying tumor characteristics. The specific tumor type and its characteristics significantly influence treatment selection. Therefore, early diagnosis and classification of brain tumors are essential for developing effective treatment strategies.

Brain tumor diagnoses and treatment methods are constantly being improved using imaging techniques to solve this problem. MRI is the most common technique used by medical experts to diagnose BT [4], and medical imaging emerged as a new technique for brain tumor treatment [5]. However, manual BT segmentation relies on the expertise of medical professionals, there is a growing need for automated methods to assist with segmentation and classification [6]. In the presence of pathology, many methods have failed for image segmentation. To overcome those issues, many researchers have developed several automatic segmentation techniques to increase the accuracy and applicability of tumor detection [7]. Over the past few years, a diverse array of brain tumor detection techniques has been developed, evolving from traditional image processing approaches to more sophisticated Neural Networks [8]. In recent years, researchers have used supervised and unsupervised, but they have typically resulted in limited segmentation accuracy [9]. Early diagnosis, surgery, and treatment depend on the accurate segmentation of brain tumors [10, 11]. However, many existing methods still struggle with low segmentation accuracy and classification of tumor regions. Therefore, accurate BT segmentation and the classification are vital for optimal treatment planning

Recently, Deep Learning (DL) models have gained significance, becoming essential tools in medical image analysis due to their ability to learn complex patterns in data [12-13]. The classification of BT in this approach utilizes a deep three-D convolutional NN to distinguish between low-grade and high-grade gliomas [14]. Segmentation tasks are performed using a convolutional neural network and a probabilistic neural network [15]. Moreover, Feature Extraction (FE) is a vital method for MRI image processing. It combines the texture, intensities, and shape-based features and classifies the image as gray, white matter, normal and abnormal areas. It efficiently selects prominent features to increase the accuracy of the diagnostic system. Furthermore, BT segmentation approaches can be classified into three categories which are based on machine learning, traditional image algorithms, and deep learning techniques. Due to its high accuracy, deep learning has become the method of choice for complex tasks. Over the past years, various Convolutional Neural Network technique has been utilized for image segmentation [16]. The classification performance depends on image features and the model [17].

In neural network-based classification, the network structure of convolution methods is complex with over-fitting and vanishing gradient problems, which minimize the automated system's performance. However, there is no powerful predictive potential associated with the existing technologies for identifying brain tumors. Most classification frameworks related to brain tumor classification face the issues of ineffective classification performance, overfitting and the inability to extract necessary features, leaving considerable room for improvement. To address these challenges, hybrid models have been suggested to enhance accuracy. To address these challenges, hybrid models have been suggested to enhance accuracy. The DCNN is designed with TL to leverage pre-trained models like AlexNet for improved feature extraction. To address the challenges of segmentation and classification, we utilize the TDA-IPH model for precise tumor segmentation, which captures the shape and structure of tumors, enhancing boundary delineation. For classification, we implement the Convolutional Transfer Learning and Visual Recurrent Learning (CTVR-EHO) model, which effectively integrates spatial and temporal dependencies in the data. Additionally, we employ EHO algorithm to optimize hyperparameters, ensuring that the model achieves its best performance by systematically exploring the hyperparameter space. This combined approach of TDA-IPH, CTVR-EHO, and EHO allows for more accurate brain tumor detection, resulting in superior diagnostic outcomes compared to existing methods.

The significant contribution of this proposed work:

- TDA- IPH is used for the segmentation of brain tumors to accurately segment small tumors from the input image.
- The AlexNet and BiVLSTM models extract low-level and high-level features from the CTVR-EHO model simultaneously. This gives importance to high-level and low-level features, so there is no loss of features during feature extraction.
- The EHO algorithm is used to tune the hyper-parameters of the AlexNet and BiVLSTM networks to achieve optimal performance.

The proposed TDA-IPH and CTVR-EHO models outperform existing approaches in terms of precision, recall, F score, accuracy, and loss.

## 2. Literature Review

In neuroimaging, noise reduction techniques are critical for enhancing the quality of MRI data, particularly in the presence of physiological artifacts such as brain, cardiac and respiratory signals. Maltbie et al. [18] investigated the use of functional MRI (fMRI) time-series entropy as a biomarker for assessing the excitatory/inhibitory (E/I) balance in the brain, employing pharmacological neuromodulation in a non-human primate model. Their research highlights the potential of fMRI entropy as a sensitive measure of neural dynamics, providing insights into brain function and the effects of neuromodulatory interventions. This work contributes to the understanding of time series affects the quality of detection in MRI. Vidaurre et al. [19] analyzed the importance of spontaneous and short-term interactions within brain networks using fMRI time-series data. Their study emphasizes how the temporal dynamics of brain activity can provide insights into neural processes affected by tumors. Specifically, identifying transient interactions reveals changes in functional connectivity that help understand a tumor's impact on surrounding structures. By analyzing time series data, researchers can enhance segmentation and classification techniques, improving differentiation between healthy and tumor-affected brain regions. This approach highlights the importance of incorporating time series dynamics in neuroimaging for more accurate brain tumor detection and classification.

Alternatively, Ferdous et al. [20] introduced LCDEiT, a data-efficient image transformer with linear complexity, designed for MRI BT classification. Their study addresses the challenges of high computational costs and data inefficiencies commonly seen in traditional transformer models. By adopting a linear complexity approach, LCDEiT significantly reduces computational demands while achieving strong performance in image classification tasks. This advancement enhances the model's capacity to effectively learn from MRI images, even with limited training data. The results also pointed out the transformer architecture in the improvement of diagnostic accuracy regarding brain tumors and relevance for the challenges of clinical neuroimaging related to images. Likewise, Shah et al. [21] introduced a novel method for classifying and localizing abnormalities in brain MRI using a channel attention-based semi-Bayesian ensemble voting mechanism combined with a convolutional autoencoder. Their approach highlights how attention mechanisms can enhance the model's focus on relvent features in MRI images, addressing common image quality issues such as noise and artifacts. By utilizing ensemble voting, the study improves decision-making through the aggregation of predictions from multiple models, while the convolutional autoencoder effectively extracts complex features from the input images. Additionally, Ghahfarrokhi and Khodadadi [22] proposed a novel approach for diagnosing human brain tumors by addressing key image-quality issues through the integration of complexity measures and texture features extracted from magnetic resonance images (MRI). Their study highlights the significance of these feature sets in enhancing image analysis for tumor detection. Complexity measures reveal underlying patterns and irregularities in the tumor, providing critical information about its structure. It helps to differentiate between healthy and abnormal tissues. By combining these features, their methodology effectively tackles challenges related to image quality and noise, resulting in improved diagnostic accuracy and reliability. This work contributes to the growing field of advanced image processing techniques aimed at enhancing tumor detection and classification in clinical practice. In addition to, Özbay and Özbay [23] proposed a brain tumor detection method that integrates interpretable feature fusion with deep hashing for high-precision MRI images. This approach prioritizes interpretability and efficiency in feature extraction by compressing MRI data through deep hashing, which retains essential tumor-related features. By fusing interpretable features, the method improves the accuracy of brain tumor detection while ensuring computational efficiency, effectively tackling the challenges posed by large MRI datasets. However, it is crucial to address potential image issues, such as noise or artifacts in MRI scans, which could affect the model's performance. Xu et al. [24] presented an advanced approach to medical image fusion. This approach merges an advanced model inspired by visual processing in the brain with artificial selection techniques and an impulse-driven neural network to effectively integrate images from various medical imaging methods. However, the study also addresses potential image issues, such as noise, artifacts, and discrepancies in modality characteristics, which could impact fusion quality and diagnostic accuracy. Therefore, an ideal preprocessing steps is needed before going for further analysis in MRI images. Deckers et al. [25] introduced an adaptive filtering method specifically designed to suppress these unwanted noise sources in MRI time series data. Their approach demonstrated significant improvements in image quality by effectively isolating and removing noise while preserving the underlying neural signals. This work underscores the importance of robust preprocessing techniques in neuroimaging and this aims directly at developing methods accurately in detecting and classifying the different types of brain tumors present from MRI scans.

For the segmentation of images, Thillaikkarasi et al. [26] used a Multiclass-Support Vector Machine (M-SVM) and Kernel-based Convolution Neural Network (K-CNN) to segment the kind of tumor as benign or malignant automatically. Thaha et al. [27] applied an Enhanced Convolutional Neural Network to identify the brain tumor using segmentation. The segmentation results are improved with a novel BAT optimization technique. The authors in [27] used M-SVM for classification and K-CNN for segmentation, a similar kind of research completed by [28]. Sajid et al. [29] implemented a hybrid model combining two and three-path CNN for brain tumor segmentation with improved results. Jia et al. [30] presented deep learning-based architecture, Fully Automatic Heterogeneous Segmentation using a Support vector machine (FAHS-SVM) to perform brain tumor segmentation. Khan et al. [31] proposed a novel cascading approach combined with handcrafted features, and CNN Model is designed for fully automatic brain tumor segmentation. The said framework is divided into two phases confidence surface modality generation and the proposed convolutional neural network approach. For feature extraction, Mittal et al. [32] developed a DL model for detecting BT segmentation. The feature is extracted by the Stationary Wavelet Transform (SWT), which divides the image and extracts the feature; New Growing Convolutional Neural Network (GCNN) is used for classification. Ozyurt et al. [33] proposed a Super Resolution Fuzzy C-means CNN (SR-FCM-CNN) model to detect the brain tumor MRI image. They used Squeeze Net and Extreme Learning Machine for feature extraction and classification. Bhargavi et al. [34] presented the CNN model designed to identify the brain tumor MRI image. Gray-Level Co-Occurrence Matrix (GLCM) and Discrete Wavelet Transform (DWT) extract the coefficients from brain tumor images to extract the features and Histo-based image segmentation. Xing et al. [35] have established Convolutional Neural Network with Element wise filter (CNN-EW) for brain networks. CNN-EW extracts hierarchical topological features. Rathore et al. [36] have proposed a neural network based on topological feature extraction, which is used to classify autism-related brain structures. Kumar et al. [37] proposed a Deep Learning model that includes global average pooling and ResNet-50 to prevent overfitting and vanishing gradient problems. Deepak and Ameer [38] introduced the CNN and SVM models for feature extraction and brain tumor classification, respectively. Kesav and Jibukumar [39] introduced the Region-Based Convolutional Neural Network (RCNN) and 2-channel CNN for brain tumor analysis. Gupta et al. [40] introduced a multi-task attention-guided encoder-decoder network (MAG-Net) to segment and classify brain tumors. Ahuja et al. [41] presented the DarkNet model for segmentation localization and classification of brain tumors. Li et al. [42] introduced a U-Net model based on LKA for the automatic segmentation of BT in MRI scans. Guan et al. [43] developed a network architecture, the Attention Guide Filter (AG) Squeeze and Excite (SE)-based VNet (AGSE-VNet), for segmenting 3D MRI brain tumor images. Ghassemi et al. [44] proposed a method utilizing GAN for multi-class classification of BTs.

Table 1. Summary of the existing approaches in BT segmentation in CNN

| Author/Year | Technique | Advantages | Disadvantages | Accuracy |
|---|---|---|---|---|
| Maltbie et al. [18], 2020 | fMRI Time-Series Entropy | Provides insights into brain E/I balance. | Limited to non-human primate models. | Precision – 98, accuracy – 98.45 |
| Vidaurre et al. [19], 2021 | Spontaneous Brain Network Interactions | Highlights the relevance of network interactions. | Complexity in interpreting transient interactions. | Accuracy – 98.78<br>Specificity – 96.7 |
| Ferdous et al. [20], 2023 | LCDEiT | Linear complexity and data efficiency for MRI classification. | Limited scalability with larger datasets. | F1 Score – 97.8<br>Accuracy – 98.11<br>Precision – 97.02<br>Recall – 96.5 |
| Shah et al. [21], 2023 | Semi-Bayesian Ensemble Voting Mechanism | Classifies and localizes abnormalities effectively. | Complexity of the ensemble method can increase training time. | Not specified |
| Ghahfarrokhi et al. [22], 2020 | Complexity Measures and Texture Features | Combines multiple features for improved diagnosis. | Feature extraction may be computationally intensive. | Accuracy – 98.9 |
| Özbay et al. [23], 2023 | Interpretable Features Fusion | Enhances interpretability of MRI data. | May require extensive preprocessing. | Not specified |
| Xu et al. [24], 2023 | Enhanced Cross-Visual Cortex Model | Improves medical image fusion. | Complexity in model design and implementation. | Not specified |
| Deckers et al. [25], 2006 | Adaptive Filter for MRI Noise Suppression | Suppresses noise effectively. | May introduce artifacts if not calibrated properly. | Accuracy - 95 |
| Thillaikkarasi et al. [26], 2019 | Kernel Based CNN and M-SVM | The tumor region is segmented correctly. | Worked on limited data samples | Accuracy - 84 |
| Thaha et al. [27], 2019 | Enhanced CNN and BAT optimization | Accurate region of brain tumor. | Different selection schemes can be adopted to improve the accuracy. | Precision-87<br>Recall -90<br>Accuracy-92 |
| Kaldera et al. [28], 2019 | CNN and Faster R-CNN | Reduced Number of Computation with good accuracy | The author proceeds only with axial plane slices. | Accuracy - 94 |
| Sajid et al. [29], 2019 | Hybrid Convolutional Neural Network | Deals with overfitting problems and Handles Data Imbalance problems | High Dropout values after every convolution layer. | Dice Score 0.86<br>Sensitivity 0.86<br>Specificity 0.91 |
| Jia et al. [30], 2020 | Automatic Heterogeneous Segmentation using Support vector machine | It eliminates the noise, which improves the quality of the tumor image. | The technique is relatively slow. | 98.51 |
| Khan et al. [31], 2020 | Cascading Approach Combined and CNN | IoT Enabled brain tumor segmentation | Accuracy Needs to improve. | Dice Similarity Scores<br>0.81 |
| Mittal et al. [32], 2019 | SWT and GCNN | A combination of SWT and GCCN improves the accuracy. | More combinations of classifiers are required to improve accuracy, but it may increase the computational cost | 98.6 |
| Ozyurt et al. [33], 2020 | Fuzzy C Means with Super Resolution, CNN, Extreme Learning Machine & Squeez Net | The high success rate of the segmentation process. | Time consumption is high. | 98.33 |
| Bhargavi et al. [34], 2019 | CNN Model, Gray Level Co-occurrence matrix & discrete wavelet transform, Histo-based image segmentation | Graphical User Interface is used | Accuracy is low | Sensitivity, Specificity<br>& Accuracy<br>60% |

| Xing et al. [35], 2018 | Convolutional Neural Network with element wise filters (CNN-EW) | Advantages of topological structure information | The Accuracy of the model is low | Accuracy 66.88<br>Sensitivity 66.44<br>Specificity 70.4 |
|---|---|---|---|---|
| Rathore A et al. [36], 2019 | Neural Network based Topological Data Analysis | Improvement in classification accuracy using a hybrid model that combines topological and correlation features | Noise in data and Heterogeneity across multiple sites is the reason for low accuracy | Accuracy 69.2% |
| Kumar et al. [37], 2021 | ResNet-50 and Global Average Pooling | Avoid overfitting and vanishing gradient issues. | The model is computationally expensive | 97.08% & 97.48% with and without data augmentation |
| Deepak and Ameer [38], 2021 | CNN and M-SVM | Less computations and memory | The additional training time required by CNN prior to feature extraction | 95.82% |
| Kesav and Jibukumar [39], 2021 | Region Based Convolutional Neural Network (RCNN) and Two Channel CNN | Reduces execution time | Limited to object detection | 98.21 |
| Gupta et al. [40], 2021 | MAG-Net | Better Feature Extraction and reduced computation | Worked on limited data samples | Precession, Recall & F1 Score is 0.98 |
| Ahuja et al. [41], 2022 | DarkNet & Super Pixel Technique | Recognize small tumors | T1 modality is classified with low accuracy. | 99.43% |
| Li et al. [42], 2022 | LKAU-Net | Spatial flexibility, long-range reliance, and local contextual information, while avoiding high computing cost and disregard of channel adaptation. | Computational complexity is high. | DC - 79.01%, 91.31%, 85.75%, and 95%<br>HD - 26.27, 4.56, and 5.87 |
| Guan et al. [43], 2022 | AGSE-VNet | Improves the accuracy and efficiency through the clever use of channel relationships, attention mechanism, full jump connections, categorical Dice loss function, and potential for clinical trials. | The model's segmentation prediction of the tumor core is slightly biased, | Dice scores of 67%, 85%, and 69%. |
| Ghassemi et al. [44], 2020 | GAN based method | Tackles the problem of low dataset size by using an unsupervised learning approach. | GAN-generated synthetic data may not always accurately represent the real data, which can affect the classification results. | Accuracy - 95.60%, Sensitivity- 94%, Specificity- 98%, Precision- 95.29%, F1- 95.10 |

## 3. Proposed TDA-IPH and CTVR-EHO Methodology for BT segmentation and classification

The framework of the proposed TDA-IPH and CTVR-EHO model carries three phases: (i) De-noising, (ii) TDA-based Segmentation, and (iii) Classification with Hyper-parameters tuning. The workflow of the proposed model is shown in Figure 1.

### 3.1. Pre-processing

In this section, we outline the pre-processing techniques applied to enhance the quality of MRI images prior to the implementation of our proposed brain tumor segmentation and classification methods. Initially, MRI images are often suffer from variations in size and orientation. To ensure the quality of the features used in our model, several rigorous pre-processing techniques were applied to the MRI images.

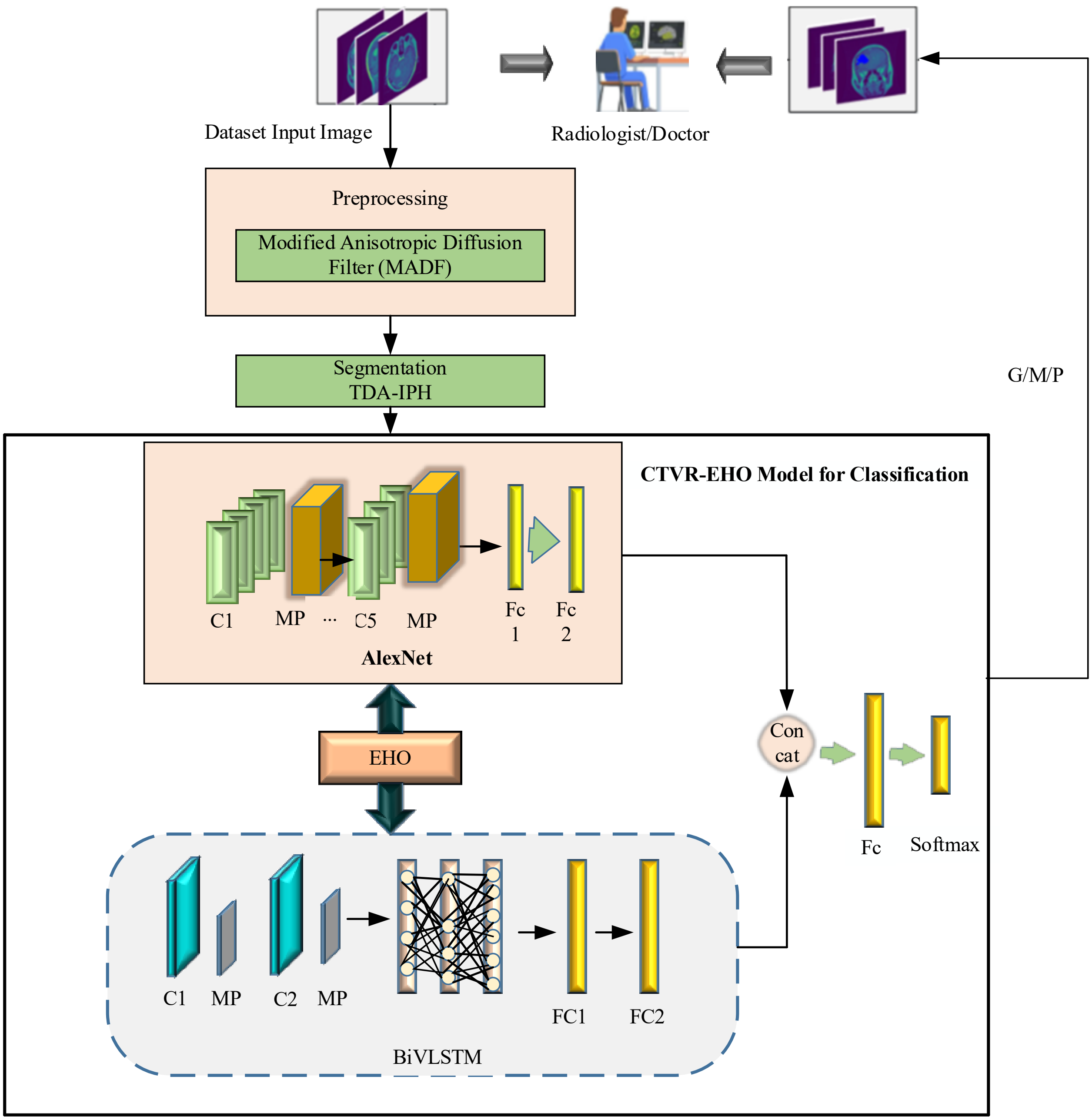

Figure 1 Architecture of proposed TDA-IPH and CTVR-EHO model

The pre-processing includes normalization, Modified Anisotropic Diffusion Filter (MADF), morphological operations, standardization. This directly impacts the model's ability to extract informative features. The pre-processing steps are given below.

3.1.1 Normalization

It adjusts the intensity values of the MRI images to a common scale, which helps in enhancing the contrast and ensuring that the model can effectively learn from the data. Normalization typically involves rescaling pixel values to a range 0 and 1[45].

3.1.2 Research design Modified Anisotropic Diffusion Filter (MADF) for De-noising

This step is critical in our method as it prepares MRI images for accurate brain tumor segmentation and classification, which would otherwise be hindered by the inherent noise present in MRI data. MRI images often contain noise due to scanning artifacts, patient movement, and equipment limitations. This noise introduces challenges in feature extraction by distorting important details such as tumor edges, texture, and intensity gradients, which are essential for accurately identifying tumor boundaries. Prior to preprocessing, these distortions result in unclear boundaries, inconsistent intensity across the image, and reduced contrast between tumor and non-tumor regions, impacting the model's ability to extract informative features.

To address these issues, we used a Modified Anisotropic Diffusion Filter (MADF) as part of our preprocessing pipeline. An isotropic diffusion filter is a commonly used noise reduction and edge preservation method. However, the anisotropic diffusion filter still generates blurring at the edge. So, we used MADF, which removes noise while preserving the edges from blurring. MADF enhances the traditional anisotropic diffusion approach by refining the threshold parameter (TP) and diffusion coefficient (DC) dynamically for each image, based on the specific content and features of that image [46]. Unlike standard diffusion methods that risk blurring edges, MADF selectively targets noise while preserving high-frequency information, particularly at the edges. It has Dynamic Adaptation of TP and DC based on gradient information, which helps retain significant edges and textures in the MRI image. Also, Iterative Adjustment of diffusion parameters, allowing more nuanced noise removal that adapts to different regions of the image.

MADF plays a key role in maintaining essential features during denoising. By preserving boundaries and enhancing contrast in regions of interest, it prevents the loss of critical structural information, such as:

**(i)** Tumor Boundaries: These are preserved sharply, allowing for clear delineation between tumor and surrounding tissue.

**(ii)** Texture and Intensity Variations: Preserving these details supports the model's ability to identify subtle variations within tumor regions that contribute to accurate classification.

In the MADF approach, TP is initially defined as the mean absolute deviation of the gradient magnitudes of the input image. However, unlike traditional methods where the threshold value remains constant during iterations, TP in MADF is adapted dynamically throughout the diffusion process. In the anisotropic diffusion model, TP plays a crucial role in noise removal and edge preservation. Although the value of TP is initially fixed, it can change in each iteration depending on the characteristics of the image at that point in the diffusion process. This adaptation allows for better preservation of edges while progressively reducing noise.

Initially, the gradient value for each pixel of a noisy input image is calculated in four directions. Then TP is estimated, and the diffusion process is applied until the required iteration. Finally, the de-noised image is obtained. In this approach, TP and DC are essential parameters for noise removal while preserving the edge from blurring. In the anisotropic diffusion model, the DC converges relatively slow, and also, in all iterations, the same threshold values create edge blurring. Therefore, TP and DC are modified in MADF. In MADF, modified DC has high convergence, and TP has different values in each iteration. DC function that converges faster than the existing models is provided in Eq. (1).

$$d = \frac{1}{1+e^{(2abs(s/t)^{2})}} + \frac{0.5}{1+e^{(2abs(s^{2}/t)}} \quad (1)$$

where, $s$ represents the gradient value calculated for each pixel of the input image and $t$ is TP. In the MADF approach, TP is adapted to be image-dependent. If only the left portion of Eq. (1) is utilized, it effectively removes noise; however, this method has a slow convergence rate and may eliminate important image features. Conversely, used the right portion allows for rapid convergence but introduces a delay in smoothing. To balance these effects, a linear combination that converges between these two terms is employed.

Additionally, the square of TP is not taken in both terms because it negatively impacts the convergence rate. The adaptation of TP enables it to adjust iteratively based on the current state of the image. It allows to effectively maintains the edges while removing the noise. Therefore, this adaptation means that, although the starting value may be fixed, the method can modify TP based on the characteristics of the current state of the image as the diffusion process proceeds. Even though TP starts as a fixed value, it can change depending on the image content, allowing for more nuanced control over the noise removal process during iterations. This helps effectively balance noise reduction and edge preservation, which would not be possible if TP were truly static throughout the filtering process.

The TP is the gradient of the input image's weighted average absolute mean deviation, which is given in Eq. (2)

$$t_{n+1} = w * Mean_{M_n}\left[\nabla M_n - Mean_{M_n}(\|\nabla M_n\|)\right] \quad (2)$$

Where Mean_(M_n )represent average operation, denotes weight. By minimizing noise and enhancing feature clarity, MADF prepares the image for segmentation. It then segments the denoised image with greater accuracy by leveraging the refined features, which were challenging to achieve in noisy images. Therefore, it ensuring that the process did not eliminate essential information and improves the clarity of tumor regions, aiding in more precise feature extraction, leading to higher accuracy in both segmentation and classification tasks

3.1.3 Morphological Operations

Following noise reduction, morphological operations (like erosion, dilation, opening, or closing) can be applied to further refine the image. These operations help in removing small artifacts, enhancing structures of interest (like tumors), and connecting disjointed parts of the tumor, making it easier to segment [47]. By applying these operations, the image quality is improved, and the accuracy of feature extraction and segmentation is enhanced

3.1.4 Standardization

Standardizing intensity values involves adjusting them so that they have a mean of zero and a variance of one. This process effectively eliminates bias in the data and ensures a uniform distribution of pixel values across all images [48]. By achieving a consistent scale, standardization enhances the suitability of the images for quantitative analysis. Moreover, it significantly boosts the effectiveness of deep learning algorithms, as these models perform optimally when input data is centered and normalized.

### 3.2. Image Segmentation using TDA-IPH

TDA is a mathematical model which utilized to extract, shape, and segment complex images. In TDA, the features are extracted by using Persistent Homology (PH). TDA is utilized in segmentation because it presents a new method of looking at patterns in data related to its structure, resulting in better segmentation results than other machine learning approaches. Initially, the collection of edge points is detected from $M$ using a wavelet-based edge detector approach. Hence, a set of $Y = y_1, y_n \ldots . y_n$ points in $M$ have been found. Then thickening of points is performed by creating a disc of radius $\beta$ with centres at each point. Larger areas of the image are covered as $\beta$ increases, and hence the disc of radius begins to overlap. Then, edges are formed by connecting the centers of the overlapping $\beta$ disc. If $\beta$ -neighbourhood of the point $r$ and $s$ contains no other points are then connected by an edge if they are within two points. The skeleton connection created in this manner is called $\beta$ skeleton. As the $\beta$ value increases, additional edges and skeleton structures are formed. The subcomplex $R_\beta$, is defined by the collection of vertices, edges, and skeletons for a given value of $\beta$. The topology created by the union of discs with radius centred at each vertex is represented by this $R_\beta$ complex. The first homology group Betti number is defined as $B_1 = D - E + 1$ for a connected graph with $D$ and $E$ being the number of edges and vertices, respectively. If we have a skeleton sub-complex in the plane, then $B_1 = O - E + D - A$ , where $O$ is the number of linked components and $A$ is the number of skeleton or faces in the complex, $E$ is number of edges in the graph, $D$ is number of vertices. The Betti number $B_1(R_\beta)$ for each sub-complex $R_\beta$ in the filtration will vary as the $\beta$change. The$B_1(R_\beta)$ for each sub-complex $R_\beta$ is given in Eq. (3).

$$B_1(R_\beta) = O - E + D - A - S \quad (3)$$

Eq. (3) defines the Betti number, which represents the relationship between key topological features of the segmentation process. This value is crucial for understanding how the structure of the graph evolves as the $\beta$ parameter changes during the segmentation. Specifically, $B_1$ influences the segmentation of porous materials by identifying different regions in the input image. To overcome challenges in segmenting porous materials from the input image, we design the Improved PH (IPH). In TDA-IPH segmentation, we use a split and merge segmentation technique to segment porous materials from input image. The Betti number plays a significant role in segmenting the image into three main regions: $P_p$-persistent, $P_t$-transient, and $d$ -skeleton, topological splitting is used to create segmentation initially.

Two parameters $\beta$ and $P$ , manage topological splitting, where $\beta$ specifies the radius of the discs and $P$ denotes persistence. Finally, it performs segmentation by merging the $P_t$-Transient, and $d$ -skeleton regions with either the $P$ -Persistent or each other. $P_p$- Persistent represent the region that are holes in $R_{\beta+P}$. On the other hand, $P_t$- Transient are smaller areas that skeleton or new edge is created in between $R_\beta$and $R_{\beta+P}$, whereas $d$ -skeleton are the smallest regions that were already skeleton structures in $R_\beta$ . The $P_t$ -Persistent areas are important because they define main objects during initial segmentation. These areas are essential to segmentation for two reasons: they are completely encircled by edges and all have a $\beta + P$radius disc. After that, these initial regions are enlarged to achieve a complete segmentation. Merging nearby areas based on their feature properties is performed to complete segmentation. In this method we utilised the average colour value for each region. Regions are merged only if $B_1(R_\beta)$ remains stable, in order to maintain the segmentation's topology. This process will reduce the parameters needed to build a PH and makes the method less flexible and more robust to reduce the noise.

| Algorithm 1: Image segmentation using TDA-IPH |
|---|
| Input: Denoised image $M$<br>Output: Segmented image |
| 1. Find the edge points $Y$ in the denoised image $M$ .<br>2. Find beta skeleton of $Y$<br>3. Compute $B_1^P(R_\beta)$ and find regions in $R^2 - R_\beta$ as $P_p$ -Persistent, $d$ -Skeleton, $P_t$ -Transient<br>4. Assume a set $A$ of regions ordered based on increasing $\beta$ values.<br>5. while ( $A \neq empty$ ) do<br>6. Let $\beta$ denote the initial face in $A$ .<br>7. Update $A$ as $A = A - \{\beta\}$<br>8. Merge $\alpha$ with the closest similar neighboring area. // $\alpha$ - 1st skeleton face in $A$<br>9. Set $B_1^P(regionmerged) = B_1^P(R_\beta)$<br>10. end while |

### 3.3. Proposed CTVR-EHO model for classification

An efficient structure to extract important information to classify the BT better is necessary since the BT is not visible against the textured background, and many small tumors are difficult to differentiate. Thus, we propose a CTVR-EHO model in which the AlexNet and Bi-VLSTM modules are parallel. The functional structure of this model is illustrated in figure 2. Here, AlexNeT extracts more apparent features like edges, corners, colour, and outlined information from images. Bi-VLSTM is used to extract high-level features from segmented BT images. Finally, extracted features from both networks are concatenated and classified. A detailed explanation of this proposed model is given in the below subsection.

#### 3.3.1 AlexNet architecture

Alexnet is one of the most efficient deep CNN designs commonly used to solve image classification issues. AlexNet demonstrated strong classification abilities, although the training took a long period. AlexNet comprises twenty-five layers, with five containing learnable weights, and the final three layers are FC layers. After the convolutional layers in the AlexNet architecture, there are ReLU, normalization, and MP layers. We had only a few hundred samples in the BT dataset, which needs to be improved to train a deep network of this size. Therefore, transfer learning (TL) is used to solve this problem, which replaces the last three layers based on our classification problem. The first five convolutional layers in this pre-trained model are transferred to extract features like edges, corners, colour, and outlined information from images. The hyperparameters in the AlexNet are tuned with the help of the EHO algorithm, which is explained in section 3.3.4.

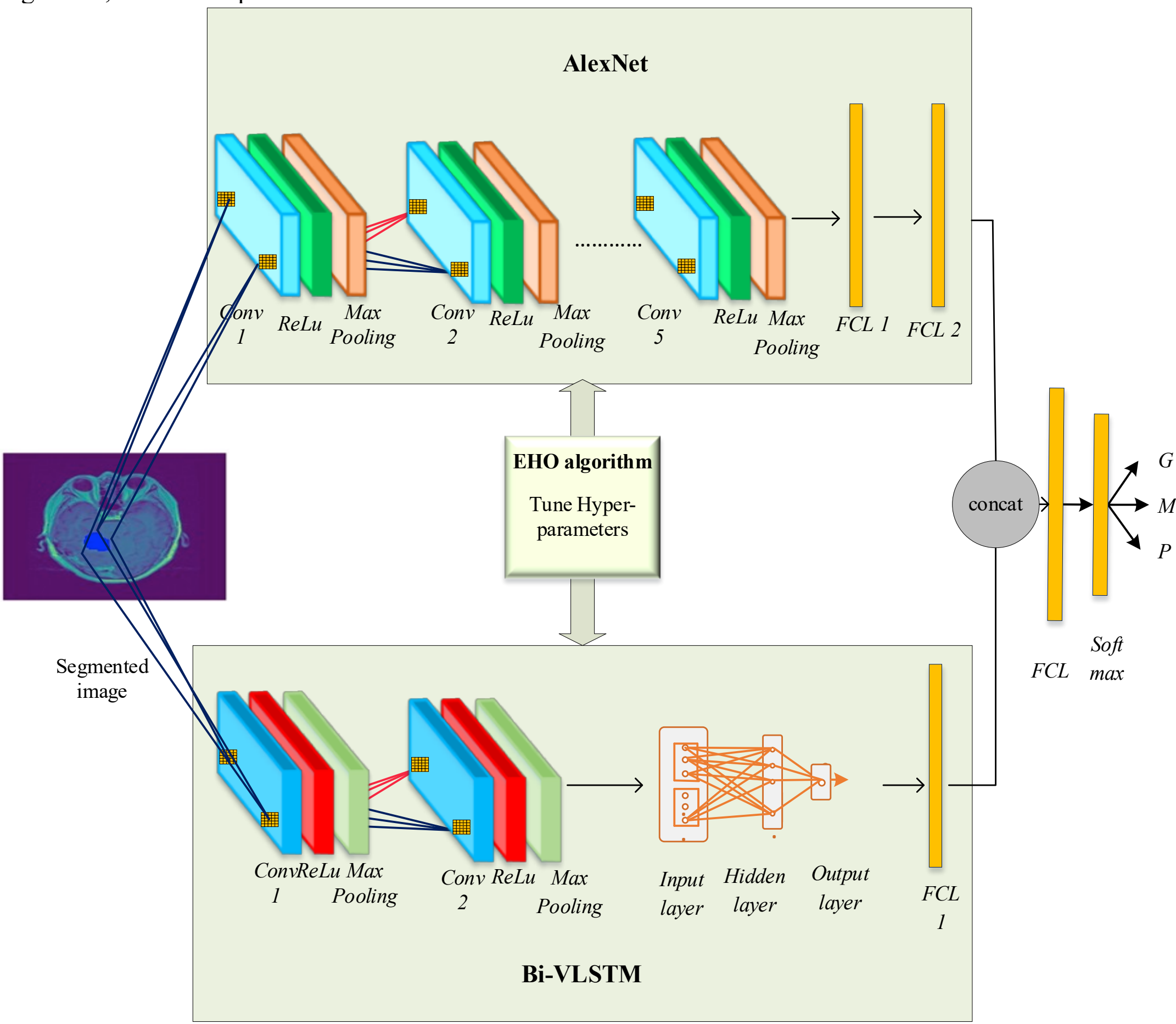

Figure 2 Architecture of CTVR-EHO for classification

### 3.3.2 Bi-VLSTM architecture

LSTM is the type of RNN that consist of the memory block. But the RNN is used for sequential and temporal data processes. Unlike LSTM, the Bi-LSTM can also store and examine the relationship among the data in two directions. Researchers have generally used a memory mechanism to solve time series issues and a visual mechanism to solve image issues. But here, we apply visual memory (VM) [39] to an image classification issue, and we fully use its information and content properties rather than the time-related feature used in existing techniques to improve the classification performance of BT. In the Bi-VLSTM model, initially, the input image's features are extracted using the convolutional layers. Max pooling (MP) layers follow each convolutional layer. The equations of the Bi-LSTM model are given below:

$$j_T = \sigma\left(W_{yj} y_T + W_{ij} i_{T-1} + W_{dj} d_{T-1} + c_j\right) \quad (4)$$

$$g_T = \sigma\left(W_{yg} y_T + W_{ig} i_{T-1} + W_{dg} d_{T-1} + c_g\right) \quad (5)$$

$$d_T = g_T d_{T-1} + j_T \tanh\left(W_{yd} y_T + W_{id} i_{T-1} + c_d\right) \quad (6)$$

$$p_T = \sigma\left(W_{yp} y_T + W_{ip} i_{T-1} + W_{dp} d_T + c_p\right) \quad (7)$$

$i_T = p_T . tanh( d_T)$ (8)

where $p$ , $j$ , $d$ and $g$ represent output gate, input gate, memory cell state, and forget gate at time $T$, respectively, $y_T$ represents input vector, $d_{T-1}$ represents last memory cell state, $W$represents weight, $i_{T-1}$denotes last hidden state, $\sigma$ represents sigmoid function, $i_T$ denotes the hidden state, and $c$ represents bias vector. In our proposed model, we use bi-LSTM model because it preserves information in both forward and backward directions. The backward hidden layer ( $h_T^c$ ) and forward hidden layer ( $h_T^g$ ) are two separate hidden layers. The input vector is taken into account by the $h_T^g$ in ascending order, i.e., $T = 1,2, \ldots, t$, and by the $h_T^c$ in descending order,

(i.e.,) $T = t-1, t-2, \ldots, 1$. The results of $h_T^c$ and $h_T^g$ are then combined to create the output $J_T$ . The implementation of this model uses the following equations:

$$i_T^g = tanh\left(W_{yi}^g y_T + W_{ii}^g i_{T-1}^g + c_i^g\right) \quad (9)$$

$$i_T^c = tanh\left(W_{yi}^c y_T + W_{ii}^c i_{T-1}^c + c_i^c\right) \quad (10)$$

$$J_T = W_{iz}^g h_T^g + W_{ii}^c h_T^c + c_z \quad (11)$$

### 3.3.3 Concatenation of features for BT classification

When training process is completed, the AlexNet model extract features like edges, corners, colour etc., from the segmented image, while simultaneously Bi-VLSTM extract high level features to aid the network in better understanding and processing of the image. After that features extracted from both AlexNet and Bi-VLSTM are concatenated for classification of BT. Concatenation of extracted features is given in Eq. (12).

$$F_C = [T_F, J_T] \quad (12)$$

where $F_C$ denotes concatenated features, $T_F$ represents features extracted using TLEA and $J_T$ represents the features extracted using Bi-VLSTM. Then, these concatenated features are given to FC layer. The output of the FC layer ( $F_y$) is given in Eq. (13).

$$F_y = f(F_c * W_w * B_b) \quad (13)$$

where $W_w$ represent weight and$B_b$represent bias of FCL. Finally, the SoftMax layer classifies the BT as glioma (G) or meningioma (M) or pituitary (P).

### 3.3.4 EHO Algorithm

Deep learning models are time-consuming and require human expertise, particularly when choosing the hyperparameter settings in the models. To save time-consuming activities and improve classification accuracy, here we use the EHO algorithm to tune the hyperparameters of Bi-VLSTM and AlexNet. The following are the steps involved in finding the optimal hyperparameters:

Step 1: Initialization

The first step is to initialize the elephant (hyper parameters) population. Hyper-parameter is initialized with a set of values which are given in Table 3 and 5.

Step 2: Fitness function

The model is trained with the hyper-parameters initialized, and then the fitness value is evaluated based on Eq. (6). Then the hyperparameters are sorted into clans according to their fitness function, and the head of the clan (matriarch) is selected based on the best fitness value. Here the accuracy is considered the criterion for evaluating a network's fitness. The fitness function is calculated using Eq. (14).

$$A_C \leftarrow \sum(x_j - k(y_j) < e_m)/K \quad (14)$$

where $K$ denotes number of inputs, $x_j$ denotes actual output for $y_j$, $k(y_j)$ denotes predicted output for $y_j$, $y_j$represents $j^{th}$input, $e_m$ denotes allowed error margin.

Step 3: Updating operation

The hyperparameter population is divided into $K$ clans and matriarch of clan $Cj$ 's, directs the movement of each hyperparameter $i$ . Each hyperparameter movement is based on the Eq. (15).

$$E_{new,Cj,i,B} = E_{Cj,i,B} + \beta(E_{best,Cj,i} - E_{Cj,i})\gamma \quad (15)$$

Where, $E_{Cj,i,B}$ and $E_{new,Cj,i,B}$ are represented the current and new position of $i^{th}$hyper parameter in clan. $\beta$ represents the scaling factor between 0 and 1. $E_{best,Cj,i}$ is the best clan position and $\gamma$ denotes the random number between 0 and 1. Fittest hyper parameter (matriarch) in each clan is updated using Eq. (16).

$$E_{new,Cj,i} = \lambda \times E_{cen,Cj} \quad (16)$$

where $\lambda$is the scaling factor and$E_{cen,Cj}$ represent the centre of the clan which is calculated using Eq. (17)

$$E_{cen,Cj} = \sum_{j=1}^{n} E_{Cj,i} / n_E \quad (17)$$

where $n_E$ is the overall number of hyper parameters in every clan.

Step 4: Separate the worst elephant

A fixed number of hyperparameters with the worst fitness function values are transferred to new position in each clan. The positions of separated worst hyperparameters is given in Eq. (18).

$$E_{worst,Cj,i} = E_{Min} + (E_{Max} - E_{Min} + 1) \times \gamma \quad (18)$$

where $E_{worst,Cj,i}$ denotes the worst fitness hyper parameter and $E_{Min} and E_{Max}$ denotes the higher and lower position of each hyper parameter.

Step 5: Convergence

After that, the same steps are repeated until reach the optimal solution of hyper parameter for classification is obtained. If the criterion status is “No”, then the step 2 to 4 will repeat until attain the convergence criteria are met.

## 4. Experimental Result and Analysis

This section evaluates the proposed model through two main sub-sections. The first sub-section (section 4.1) outlines the experimental setup, dataset description, and details the evaluation metrics used. The second sub-section (section 4.2) presents the Sensitivity Analysis of EHO Parameters, (section 4.3) presents Qualitative analysis and (section 4.4) provides Evaluation Metrics and (section 4.5) given

Comparative analysis as comparisons with other existing methods on MRI segmentation and classification.

## 4.1 Experimental Setup

The experiments are conducted on PYTHON 2021.1.3 version, i3 core, and 8GB RAM, equipped with an Intel i3 processor and a storage capacity of 256 GB. The environment operates within a Jupyter notebook, which provides an interactive platform for coding and visualization. The proposed approach is implemented using Python 2021.1.3, leveraging essential libraries such as Scikit-learn, Keras, TensorFlow, Seaborn, Matplotlib, NumPy, and Pandas.

### 4.1.1 Data Spilt

The validation strategy outlined is essential for evaluating a model's performance in machine learning. It begins with a 70%-30% train-test split, where 70% of the dataset is allocated for training and 30% for testing, ensuring the model learns from a substantial portion of the data while retaining enough unseen data for assessing generalization. In our study, we chose a 70/30 split for the training and testing data to balance model training and evaluation effectively. This ratio is widely used in machine learning due to its simplicity and effectiveness, serving as a common heuristic rather than being based on a specific prior paper. With 70% of the data is allocated for the training, we ensured that the model had enough examples to learn the underlying patterns and features necessary for accurate brain tumor detection, particularly important given the variability in tumor appearances across different cases. The remaining 30% serves as an independent test set to evaluate the model's performance, helping to assess its generalization capability and ensuring that the results are not biased toward the training data. This ratio helps to reduce overfitting by splitting the data into training and testing sets, thereby preventing the model from memorizing the training data. It also evaluates generalization by testing model performance on unseen data, which is crucial for real-world applications.

Additionally, the 70/30 split strikes a balance in the bias-variance tradeoff, minimizing the risk of underfitting and overfitting. Our confidence in this testing setup is further reinforced by employing cross-validation techniques, ensuring the model's performance remains stable across various data subsets. We also monitored performance metrics like accuracy, precision, and recall to validate the reliability of our predictions. Although we used a single split, shuffling the dataset and testing different splits could yield similar outcomes; however, maintaining a representative sample in both the training and testing sets is essential, as variability in the data, such as differences in tumor types or imaging conditions, may affect the results.

### 4.1.2 5-fold cross validation

To further enhance model robustness, 5-fold cross-validation is employed, dividing the training dataset into five subsets. The model is trained on four folds and validated on the remaining one, repeating this process five times to gauge generalization and mitigate overfitting by testing the model across multiple data partitions. Additionally, early stopping is implemented to monitor validation loss during training, halting the process when no improvement is observed for a specified number of epochs, which prevents overfitting by avoiding unnecessary training on noise. Finally, the model's performance is comprehensively evaluated using metrics such as precision (accuracy of positive predictions), recall (ability to identify all relevant instances), F1-score (harmonic mean of precision and recall), and accuracy (overall correct predictions), providing a thorough assessment of its sensitivity and specificity in segmentation and classification tasks.
The results in Table 2 indicate that the proposed model achieves exceptional performance across all evaluated metrics. With a precision of 98.7% to 99.2%, the model demonstrates a high level of accuracy in its positive predictions. The recall of 98.1% to 98.94% indicates that the model successfully identifies a vast majority of actual positive instances. The F1-score of 99.3% to 99.4% reflects a strong balance between precision and recall, suggesting that the model is adept at minimizing false positives and negatives. Furthermore, the accuracy of 98.9% to 99.6% highlights the model's overall effectiveness in correctly classifying instances. The low validation loss of 0.011 reinforces the model's capability to generalize well, indicating minimal overfitting during training. Overall, these results affirm the robustness and reliability of the proposed TDA-IPH model in segmentation and classification tasks.

Table 2: 5-fold Cross-Validation Performance Metrics

| Iteration | Testing set | Accuracy | Precision | Recall | F1-score |
|---|---|---|---|---|---|
| 1 | Fold 1 | 98.9 | 98.7 | 98.1 | 99.3 |
| 2 | Fold 2 | 99.0 | 99.1 | 98.9 | 98.7 |
| 3 | Fold 3 | 98.9 | 99.1 | 99.0 | 99.8 |
| 4 | Fold 4 | 99.4 | 98.9 | 99.91 | 99.5 |
| 5 | Fold 5 | 99.6 | 99.2 | 98.94 | 99.4 |

### 4.1.3Parameters and Hyperparameters description

The proposed model parameters and hyperparameters details are given below;
Table 3 presents the hyperparameters and their corresponding range of values for AlexNet, a convolutional NN consists of three FCL and 5 convolutional layers, which is utilized for classification. The table specifies the filter sizes in the convolutional layers, which can vary between 3, 5, and 7 to capture different features in the input images. For pooling, filter sizes of 2 or 3 are used, with the choice between Max Pooling and Average Pooling (MP, AVG) for down-sampling. The fully connected layers include neurons with sizes from 128 to 512 to control feature depth, while the activation functions introduce non-linearity to enhance learning capacity. Dropout rates of 0.3, 0.4, and 0.5 are applied for regularization, helping to reduce overfitting by randomly deactivating a fraction of neurons during training.

Table 3. Hyper parameters of AlexNet and their range of values

| Layers | Hyper parameters | Values range |
|---|---|---|
| Convolutional | Size of filter | 3,5,7 |
| Pooling | Size of filter<br>Type of pooling | 2, 3<br>MP, AVG pooling |
| FCL | Neurons | 128, 256, 512 |
| Learning | Activation function | Leaky ReLU, Elu, ReLU |
| Regularization | Dropout | .3, .4, .5 |

Table 4 describes the parameters of the Elephant Herding Optimization (EHO) algorithm, which is applied to tune the hyperparameters of both AlexNet and BiVLSTM for optimal performance in classifying MRI images of brain tumors. The parameters include a population size of 10, divided into 12 clans to simulate a multi-solution search environment. The beta and gamma values, set at 0.5 and 0.1, respectively, are crucial for updating the search strategy and influencing the exploration-exploitation balance during optimization. The generation index, which keeps track of the optimization process's progress, is crucial for monitoring convergence towards optimal hyperparameters. This tuning with EHO aims to refine the models' accuracy and generalizability, enhancing their effectiveness in brain tumor classification based on MRI features.

Table 4: Parameters of EHO algorithm

| Parameters | Value |
|---|---|
| Population size | 10 |
| Number of clan | 12 |
| Beta | 0.5 |
| Gamma | 0.1 |
| Generation index | 1 |

Table 5 describes the hyperparameters and value ranges for Bi-VLSTM, a bidirectional long short-term memory model used for sequence classification tasks. The BiVLSTM model includes hyperparameters such as the number of hidden layers, which can range from 0 to 2, and the number of neurons, which can vary from 20 to 200 to control the model's complexity and depth. The learning rate for training ranges from 0.005 to 0.2, allowing adjustment of model convergence speed, while dropout values range from 0 to 1 to introduce regularization by randomly deactivating neurons. Batch sizes range from 1 to 512, giving control over the amount of data processed in each training iteration, thereby affecting memory usage and convergence stability.

Table 5. Hyper parameters of BiVLSTM and their range of values

| Hyper parameters | Values range |
|---|---|
| Hidden layers | 0 to 2 |
| Neurons | 20 to 200 |
| Learning rate | .005 to .2 |
| Dropout | 0 to 1 |
| Batch size | 1 to 512 |

#### 4.1.4 **Dataset Description**

This section provides the simulation result and comparative analysis of MRI classification using the proposed and existing method. The Figshare brain dataset is used in this work. The dataset contains 3064 T1 weighted contrast-enhanced MRI images, of which 708 images are Meningioma (M), 1426 images are G3.1ioma (G), and 930 images are Pituitary (P) tumors. The slices are collected from 233 patients. It contains three kinds of brain tumors: meningioma, pituitary, and glioma. From the dataset, 70% of images are used for training, and the remaining 30% are used for testing. To know more about the dataset, refer [49], and the authors in [50,51] also used the figshare dataset in their research.

Table 6: summarizes the distribution of MRI images within the Figshare brain dataset, which comprises a total of T1-weighted contrast-enhanced MRI images sourced from 233 patients. The dataset is categorized into three types of brain tumors: Meningioma, Glioma, and Pituitary tumors. The total number of images for each tumor type is listed below. The training and testing distribution for each tumor type is also detailed, highlighting the number of images used in each category. This structured approach enables a clear understanding of the dataset's composition, crucial for evaluating the effectiveness of the proposed and existing classification methods.

Table 6: training and testing details of Figshare dataset

| Dataset | | Total | Training | Testing |
|---|---|---|---|---|
| Figshare | Glioma | 1426 | 1144 | 282 |
| | Meningioma | 708 | 535 | 173 |
| | Pituitary | 930 | 729 | 201 |

Algorithm 2: EHO algorithm for CTVR hyper parameter tuning

1. Initialization: Set the generation counter $s$ to 1; initialize population of elephant (hyper parameter); set maximum generation $mxg$ value.
2. Find fitness function
3. while ( $s \leq mxg$ ) do
4. Sort the hyper parameters according to their fitness value.
5. for all $Cj$ in hyper parameter population do
6. for all hyper parameters in $Cj$ do
7. Update $E_{Cj,i,B}$ and generate $E_{new,Cj,i,B}$ by Eq. (15).
8. if $E_{Cj,i,B} = E_{best,Cj,i}$ then
9. Update $E_{Cj,i,B}$ and generate $E_{new,Cj,i}$ by Eq. (16).
10. end if
11. end for
12. end for $Cj$
13. for all clans $Cj$ in the population do
15. Separate the worst hyper parameter by Eq. (18).
16. end for
17. Population is evaluated by the positions updated.
18. $s = s + 1$
19. end while
20. return the optimal solution.

### 4.2 Sensitivity Analysis of EHO Parameters

In optimization algorithms, the selection of parameters significantly influences performance, affecting both the speed of convergence and the quality of solutions. The EHO algorithm, an optimization method serves as a critical tool for understanding how variations in parameter settings impact the behavior and outcomes of the algorithm. By systematically adjusting parameters, we can identify the most effective configurations that yield optimal results. In this analysis, examine several key parameters integral to the EHO algorithm: population size, number of clans, beta (β), gamma (γ), and convergence threshold were chosen based on extensive experimental analysis. Each of these parameters plays a distinct role in shaping the search dynamics, influencing how well the algorithm performs. Understanding the sensitivity of these parameters is essential for optimizing the EHO's performance, ensuring that it converges efficiently while maintaining high accuracy. The table 7 explores the effects of various parameter settings on the performance of the EHO algorithm, particularly examining convergence speed and accuracy. Each parameter as population size, number of clans, beta ($\beta$), gamma ($\gamma$), and convergence threshold plays a distinct role in balancing exploration and exploitation in the search space. Moderate values across these parameters, specifically a population size of 10, 12 clans, $\beta$ = 0.5, $\gamma$ = 0.1, and a convergence threshold of 0.01, produced the highest accuracy 99.2% and converged efficiently, suggesting that these settings achieve an optimal balance. In particular, adjusting the population size and clan count revealed that larger values enhanced accuracy slightly, though with diminishing returns, while moderate $\beta$ and $\gamma$ values offered the best blend of precision and diversity in solutions. Furthermore, a middle-range convergence threshold ensured both stability and efficiency in convergence, achieving satisfactory results without excessive computational load. This analysis demonstrates that fine-tuning each parameter can significantly improve the effectiveness of the EHO algorithm, highlighting the value of balanced parameter settings to optimize both accuracy and speed.

Table 7: Impact of EHO Parameter Settings on Convergence Speed and Solution Accuracy

| Population size | Number of clans | Beta $\beta$ | Gamma $\gamma$ | Convergence Threshold | Iterations of converge | Accuracy |
|---|---|---|---|---|---|---|
| 5 | 8 | .2 | 0.01 | 0.01 | 120 | 99.06 |
| 10 | 12 | .5 | 0.1 | 0.001 | 150 | 99.2 |
| 15 | 16 | .8 | 0.03 | 0.0001 | 100 | 98.89 |

### 4.3 Qualitative Analysis

Axial, sagittal, and coronal views of three different types of tumors from the figshare dataset are shown in Figure 3. Input

MRI images from the dataset are 2-D images with a size of 512*512 pixels. This work initially sends images to the MADF for noise reduction. Then the pre-processed BT image is segmented and classified using the TDA-IPH and CTVR-EHO models. The segmented images and classifier outputs are shown in Figure 4.

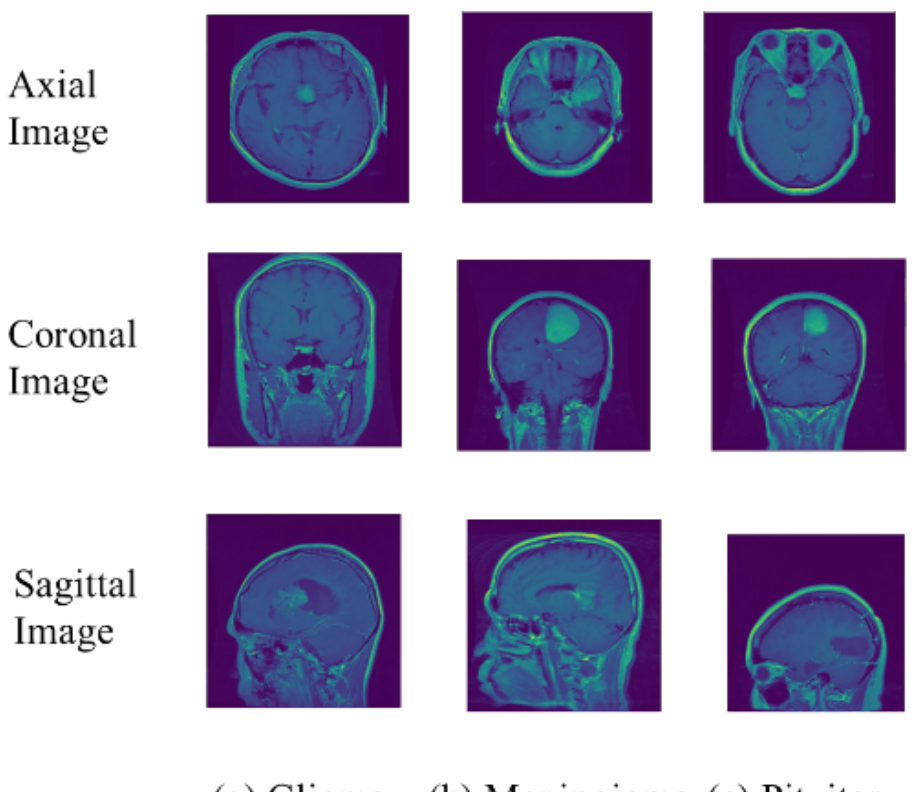

Figure 3 Sample BT image from dataset

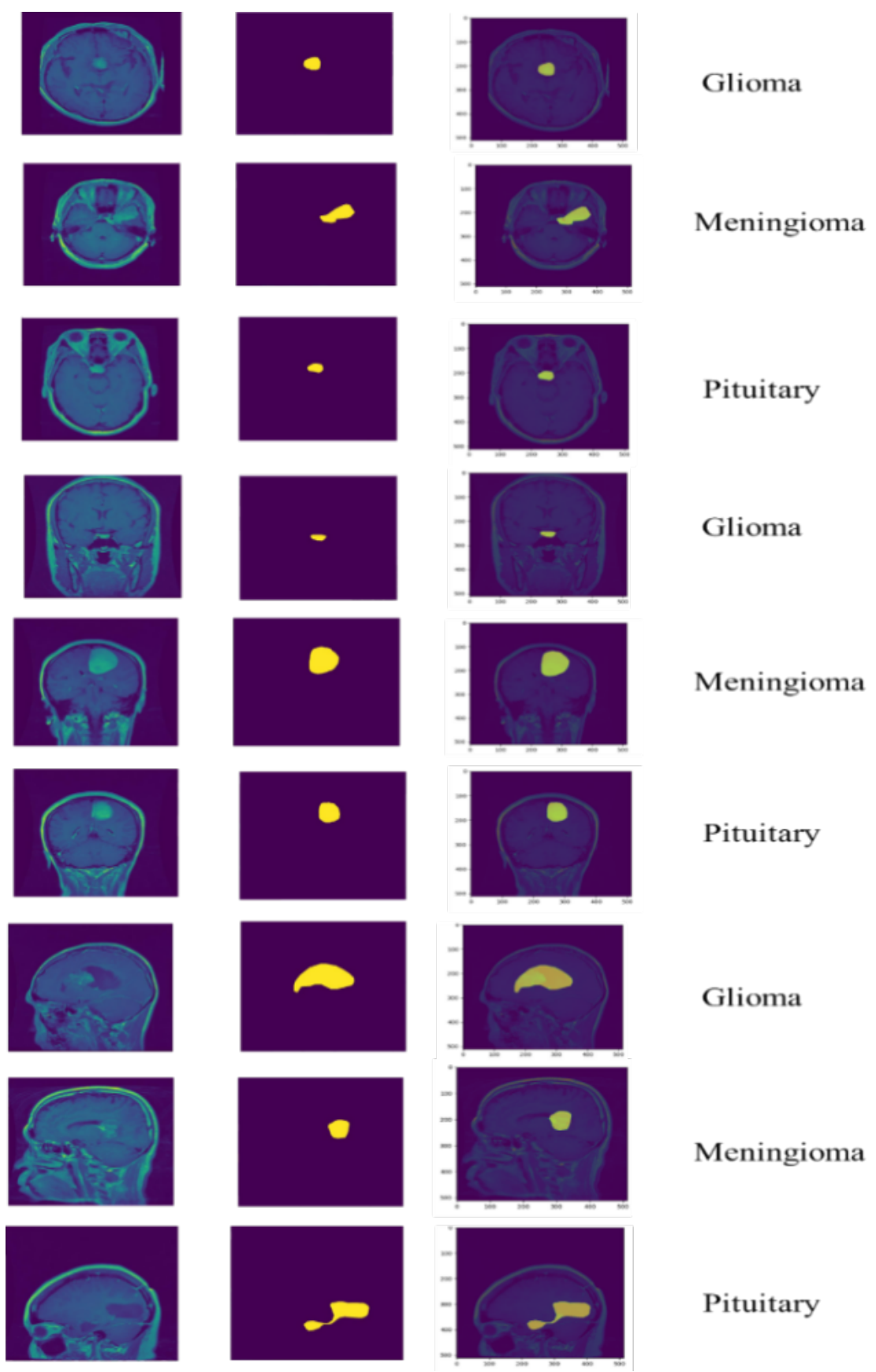

Figure 4 Outcome from TDA-IPH and CTVR-EHO

### 4.4 Evaluation Metrics

The proposed TDA-IPH and CTVR-EHO model performance is measured by applying the BT dataset, and the analysis is compared with existing methods in terms of precision, accuracy, recall, loss, and F score. A confusion matrix is the measure for determining how effectively the machine learning algorithm categorizes the given information. There are four combinations of the predicted and actual values, such as False Positive (FP), True Positive (TP), False Negative (FN), and True Negative (TN).

Our primary objective in this research is the segmentation and classification of BT. In the segmentation phase, the model distinguished between tumor and non-tumor regions. For classification, the model differentiated among glioma tumors, meningioma tumors, and pituitary tumors. To assess model performance, we constructed a confusion matrix to calculate true positives, false positives, true negatives, and false negatives, in which the FP and FN is critical metrics for identification. These metrics serve as foundational elements for assessing how well our model differentiates between tumor and non-tumor cases. The implications of these classifications are significant, as they directly impact patient care and treatment decisions. A false positive occurs when the model incorrectly predicts the presence of a tumor in a patient who does not have one. This misclassification can lead to unnecessary anxiety and medical interventions, such as biopsies or surgeries, which can cause distress and additional health risks to patients. Therefore, minimizing the rate of false positives is vital to avoid overtreatment and ensure that patients receive appropriate care tailored to their actual health status. Conversely, a false negative occurs when the model fails to identify an existing tumor, resulting in a prediction that there is no tumor present. This misclassification can have severe consequences, potentially delaying critical treatment for patients with actual tumors. The failure to detect a tumor can lead to disease progression, significantly affecting patient outcomes. Thus, accurately calculating and understanding both FP and FN is essential in clinical settings where patient health is at stake. To derive FP and FN from the confusion matrix, we can define these metrics as follows: True Positives represent correctly identified tumor cases, while True Negatives denote correctly identified non-tumor cases. False Positives are defined as the incorrectly identified tumor cases (predicted tumor but actual non-tumor), and False Negatives are the missed tumor cases (predicted non-tumor but actual tumor). These metrics have provided insights into accurately identifying affected images and correctly detecting and classifying tumor presence.

Formulas used to calculate performance metrics are given below.

The Fp calculated by identifying instances where the model predicted a tumor but the actual label is non-tumor.

FP=Total Predicted Tumor−True Positives

$$FP = Total\ Predicted\ Tumor - True\ Positive \tag{19}$$

The FN by counting the actual tumor pixels that were incorrectly classified as non-tumor.

$$FN = Actual\ Tumor - True\ Positive \tag{20}$$

These calculations provide a clear representation of the model's performance, allowing us to quantify how often our model correctly identifies tumors versus how often it misclassifies healthy tissue as tumors or fails to detect actual tumors.

Eq. (21) is used to calculate precision ($Pr$).

$$Pr = \frac{tp}{fp+tp} \tag{21}$$

Eq. (22) is used to calculate recall ($Re$).

$$Re = \frac{tn}{fn+tn} \tag{22}$$

Eq. (23) is used to calculate Accuracy ($Ac$).

$$Ac = \frac{tn+tp}{fn+fp+tn+tp} \tag{23}$$

Eq. (22) is used to calculate F score ($F-1S$).

$$F-1S = \frac{2\times Re \times precision}{Re + precision} \tag{24}$$

where $tp$ denotes the true positive, $tn$ denotes true negative, $fp$ denotes false positive and $fn$ denotes false negative.

#### 4.4.1 **Confusion Matrix**

A confusion matrix, also known as an error matrix, is a structured table that presents information on actual labels (ground truth) and the model's predicted classifications. It provides a comprehensive summary of model performance, offering detailed insights into how effectively the model generalizes across each class. Typically, the matrix displays ground truth values along the y-axis, while predicted class labels are arranged along the x-axis.

The confusion matrix serves as a crucial tool in the evaluation of brain tumor segmentation and classification, particularly when analyzing the performance of deep learning models applied to datasets. The confusion matrix offers an in-depth analysis of the model's classification results, enabling a thorough evaluation of its accuracy in correctly identifying and categorizing tumor regions. By illustrating true positive, true negative, false positive, and false negative predictions for each tumor and non-tumor region, including glioma, meningioma, and pituitary, the confusion matrix enhances the interpretability of the model's performance metrics.

In brain tumor segmentation, a high number of true positives coupled with minimal false positives and false negatives signifies the model's effectiveness in accurately delineating tumor boundaries, which is critical for subsequent treatment planning and intervention. Conversely, in tumor classification, the confusion matrix enables the identification of specific classes where the model may struggle, such as the presence of false negatives in certain tumor types, which can have significant clinical implications. Understanding these nuances allows researchers and practitioners to refine their models, focusing on areas that require improvement to enhance diagnostic reliability.

Figure 5 illustrates the segmentation confusion matrix which is used to measure how well the model differentiates between tumor and non-tumor regions in MRI images. The segmentation of brain tumors using deep learning techniques is crucial for accurately identifying and delineating tumor regions from surrounding healthy tissue in medical imaging. In this context, the confusion matrix reveals a highly effective performance of the model, showcasing 180 true positive cases, indicating that the model successfully identified all actual tumor instances. True Negatives (TN) indicate pixels accurately identified as non-tumor, while False Positives (FP) represent non-tumor pixels incorrectly classified as tumor. The model's low False Positive rate suggests minimal misclassification of non-tumor regions as tumor, a crucial factor in maintaining high segmentation precision. With only 2 false positives, it demonstrates a strong ability to minimize incorrect tumor detection, thereby reducing the risk of unnecessary interventions. Furthermore, the absence of false negatives signifies that the model did not overlook any tumors, which is critical in ensuring timely diagnosis and treatment. The true negatives of 90 further highlight the model's reliability in accurately classifying healthy tissue. False Negatives (FN), where actual tumor pixels were incorrectly classified as non-tumor, also remained low, resulting in a high recall score of 98.94%. This score highlights the model's ability to detect nearly all true tumor regions, ensuring minimal oversight of critical tumor boundaries. Overall, the segmentation confusion matrix confirms that the model effectively outlines tumor regions with a balanced precision (99.2%) and recall (98.94%), resulting in an F1-score of 99.4%. These high values signify the model's robustness in tumor delineation, a critical aspect for accurate surgical planning and treatment.

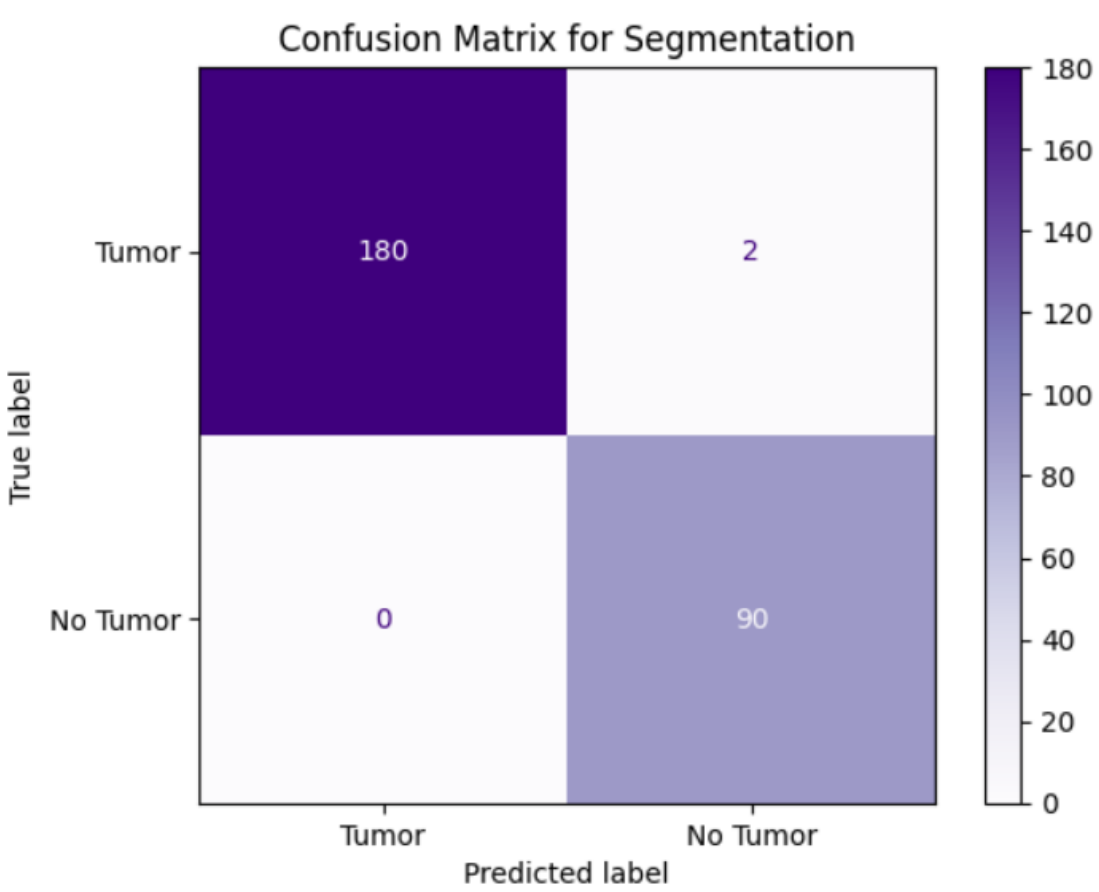

Figure 5 Confusion matrix for brain tumor segmentation

Figure 6 represents classification confusion matrix evaluates the model's performance in distinguishing among three types of brain tumors: glioma, meningioma, and pituitary tumors. Each row in this matrix represents actual tumor classes, while each column denotes the predicted classes, with the diagonal values indicating correctly classified instances. The confusion matrix for classification demonstrates solid performance across the different tumor types, with true positive counts of 75 for Glioma, 70 for Meningioma, and 55 for Pituitary tumors. Notably, the absence of false positives across all classes indicates a high level of accuracy in distinguishing between tumor types and healthy tissue. However, the detection of two false negatives in the Meningioma category raises concerns, as these missed cases could impact patient management and treatment decisions. The model demonstrates its potential as a valuable tool in clinical settings by accurately identifying most tumors and achieving zero false positive rates. Nonetheless, addressing the instances of misclassification, particularly in the Meningioma category, is essential for enhancing the model's overall reliability and ensuring comprehensive patient care in the context of brain tumor classification. For example, the recall rates for each class are similarly high: glioma at 99.6%, meningioma at 98%, and pituitary at 99.6%, highlighting the model's consistent detection capability across tumor types. This precision across classes emphasizes the model's accuracy in distinguishing tumor types, a critical factor in ensuring appropriate and specific treatment plans for each tumor type.

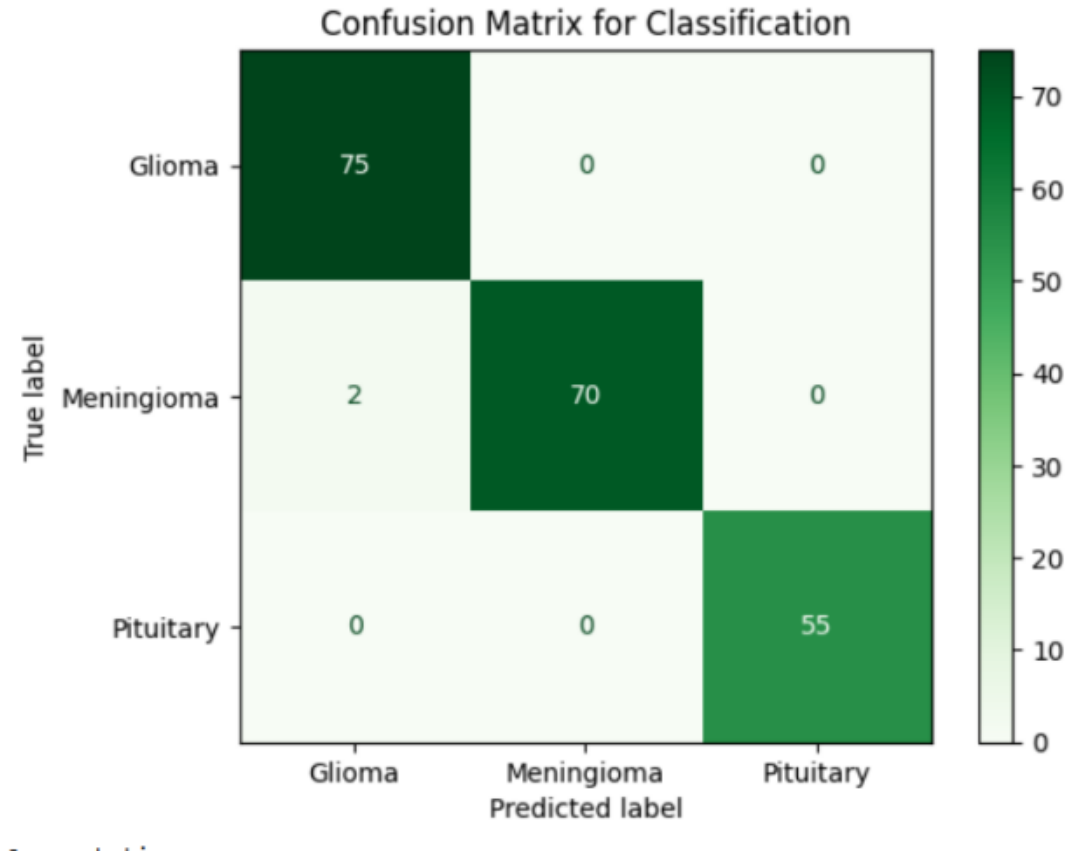

Figure 6: Confusion matrix for brain tumor classification

In medical imaging for brain tumor analysis, False Positives and False Negatives are critical for assessing a model's reliability, both in segmentation and classification tasks. In segmentation, FP errors occur when healthy tissue is mistakenly identified as a tumor, which could lead to unnecessary procedures, anxiety, and increased healthcare costs. Our model's low FP rate, with only two cases of misclassified non-tumor regions, demonstrates its precision in avoiding such errors. Conversely, FN errors in segmentation, where actual tumor regions are missed, pose a significant risk by potentially delaying diagnosis and treatment. Our model achieves a high recall of 98.94%, indicating that nearly all true tumor areas are correctly detected, thereby enhancing the reliability of tumor boundary delineation essential for treatment planning. In classification, FPs represent cases where one tumor type is misclassified as another, but the model's zero FP rate here underscores its high specificity, which is crucial for selecting appropriate, individualized treatment and avoiding misinformed interventions. FN errors in classification, though minimal, include two instances where meningiomas were missed, highlighting the importance of accuracy to prevent misdiagnosis. Together, the model's low FP and FN rates across both tasks demonstrate a high level of reliability, making it a strong candidate for accurate and effective medical diagnostics, ultimately supporting better patient care.

**4.5 Comparisons analysis**

This section presents the detailed comparative analysis of performance of the proposed models, TDA-IPH for segmentation and CTVR-EHO for classification, against several established methodologies in the field of brain tumor detection and classification. This analysis highlights the effectiveness of our models in accurately segmenting and classifying tumor regions from MRI scans.

Table 8 presents a comparision analysis of segmentation results across various models, focusing on their performance metrics, including precision, recall, F score, accuracy, and loss. The results indicate that the proposed TDA-IPH model significantly outperforms established segmentation techniques such as wU-Net [52], UNet [53], and Unet++ [54]. With a precision of 99.2%, a recall of 98.94%, an F score of 99.4%, and an accuracy of 99.6%, TDA-IPH demonstrates superior capability in accurately identifying tumor regions, even those that are small or difficult to detect. The reduction in loss to 0.011% further underscores the model's effectiveness, making it a robust option for medical image segmentation compared to its counterparts.

Figure 7 provides a visual representation of the segmentation results achieved by the various models, illustrating the effectiveness of the proposed TDA-IPH method. The figure likely depicts side-by-side comparisons of segmented images, showing how well each model identifies tumor boundaries in MRI scans. The visual analysis reinforces the quantitative findings from Table 9, highlighting TDA-IPH's ability to delineate tumors with higher accuracy and clarity than traditional methods. This figure serves as a crucial complement to the table, enabling readers to visually assess the segmentation quality and the model's performance in identifying critical features in medical images.

Table 9 details the classification performance of the proposed CTVR-EHO model alongside several existing models, including ResNet-50 [55], VGG16 [56], and VGG19 [57]. Each model's performance is evaluated based on precision, recall, F score, accuracy, and loss for different tumor classes (G, M, P). The proposed CTVR-EHO model achieves outstanding results with a precision of 99%, recall of 99.6%, F score of 99.4%, and accuracy of 99.2%, outperforming the existing networks in all categories. The relatively low loss of 0.9% further indicates the model's robustness and reliability in classifying brain tumors, thereby confirming its efficacy as an advanced solution for tumor classification tasks in medical imaging.

Figure 8 illustrates the results of the classification process, providing a graphical analysis of the performance metrics obtained from various models. The figure likely includes charts or plots that depict the precision, recall, F scores, accuracy, and loss of each model, making it easy to compare the proposed CTVR-EHO model with existing classification techniques. By visualizing these results, Figure 8 effectively summarizes the findings from Table 10, allowing for a straightforward assessment of how the CTVR-EHO model excels in accurately classifying different types of brain tumors. This visual representation enhances the reader's understanding of the model's capabilities and performance, supporting the quantitative data provided in the table

Table 8. Comparison of segmentation results

| Segmentation models | Precision | Recall | F score | Accuracy | Loss |
|---|---|---|---|---|---|
| wU-Net [52] | 84% | 96% | 98% | 94% | .030% |
| UNet [53] | 97% | 95.4% | 97.6% | 95 % | .024% |
| Unet++ [54] | 95% | 97.34% | 98.9% | 98% | .021% |
| Proposed TDA-IPH | 99.2% | 98.94% | 99.4% | 99.6% | .011% |

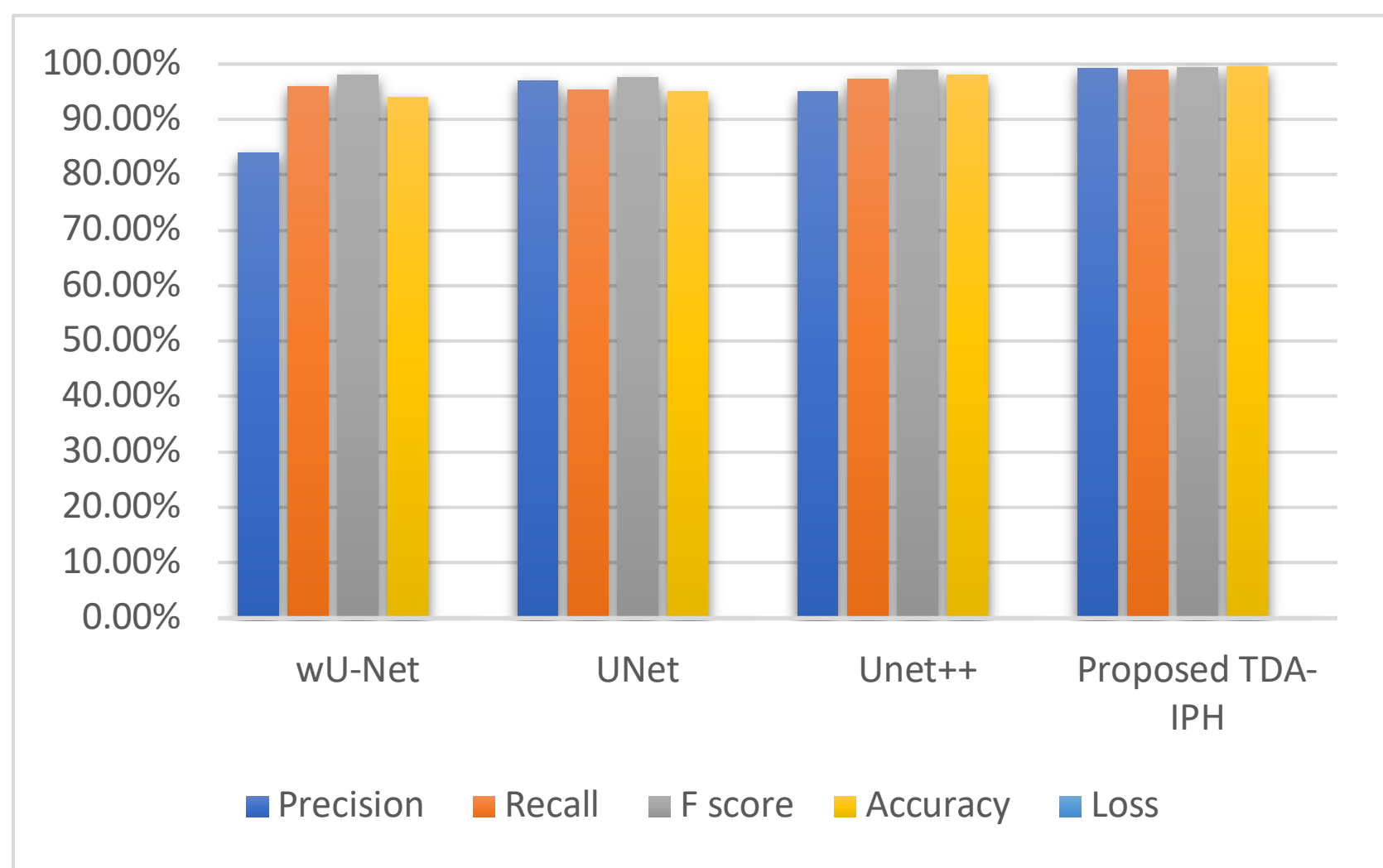

Figure 7: Segmentation result analysis

Table 9. Comparison of classification results

| Classification models | class | Precision | Recall | F score | Accuracy | Loss |
|---|---|---|---|---|---|---|
| ResNet-50 [55] | G<br>M<br>P | 93%<br>88%<br>95% | 99%<br>87%<br>99% | 94%<br>87%<br>97% | 98.14% | .31% |
| VGG16 [56] | G<br>M<br>P | 91%<br>90%<br>98% | 85%<br>80%<br>99% | 93%<br>85%<br>99% | 96% | .26% |
| VGG19 [57] | G<br>M<br>P | 93%<br>91%<br>96% | 96%<br>80%<br>99% | 94%<br>85%<br>98% | 98.9% | .25% |
| Proposed CTVR-EHO model | G<br>M<br>P | 99%<br>99%<br>98% | 99.6%<br>98%<br>99.6% | 98.7%<br>98%<br>99.4% | 99.2% | .9% |

Table 10 provides a comprehensive comparison of various state-of-the-art (SOTA) methods for brain tumor segmentation and classification in MRI, showcasing key performance metrics, including precision, recall, F-score, and accuracy. The table includes a wide array of approaches, ranging from traditional CNN [27] and advanced frameworks like Faster R-CNN [28] to models incorporating specialized methods such as handcrafted features, fuzzy clustering, element-wise filters, and GAN pre-training. Each technique is evaluated on datasets such as BRATS, Figshare, and other custom or IoT-enabled datasets, highlighting the variety of data sources used for model training and validation. The proposed TDA-IPH and CTVR-EHO model demonstrates superior performance over existing methods, achieving a precision of 99.67%, recall of 99.23%, F-score of 99.59%, and accuracy of 99.8%. Compared to other high-performing models, such as LKAU-Net [42], AGSE-VNet [43] and CNN-SVM [38], the proposed model excels in all metrics, showcasing its robustness and effectiveness in accurately segmenting and classifying brain tumors. This superior performance suggests the model's potential for clinical applications, where precise segmentation and classification are crucial for patient care. The table provides a detailed benchmarking of the proposed method against numerous SOTA models, highlighting its improvements and contributions to the analysis of medical image.

Figure 9 visually compares the performance of the proposed TDA-IPH and CTVR-EHO model with a selection of several existing SOTA BT methods for classification and segmentation. A subset of techniques from Table 10 is chosen to illustrate the performance contrast, with metrics such as precision, recall, and accuracy plotted for comparison. This figure highlights the proposed model's enhanced performance across multiple metrics, showcasing its precision, recall, and accuracy as compared to randomly selected, prominent models in the field. The graphical representation offers a clear understanding of how the proposed model consistently outperforms others in accuracy and robustness, reinforcing its suitability for real-world clinical applications. The random selection of competing techniques for the figure allows for a focused comparison, whereas the complete details are provided in Table 10 for thorough analysis.

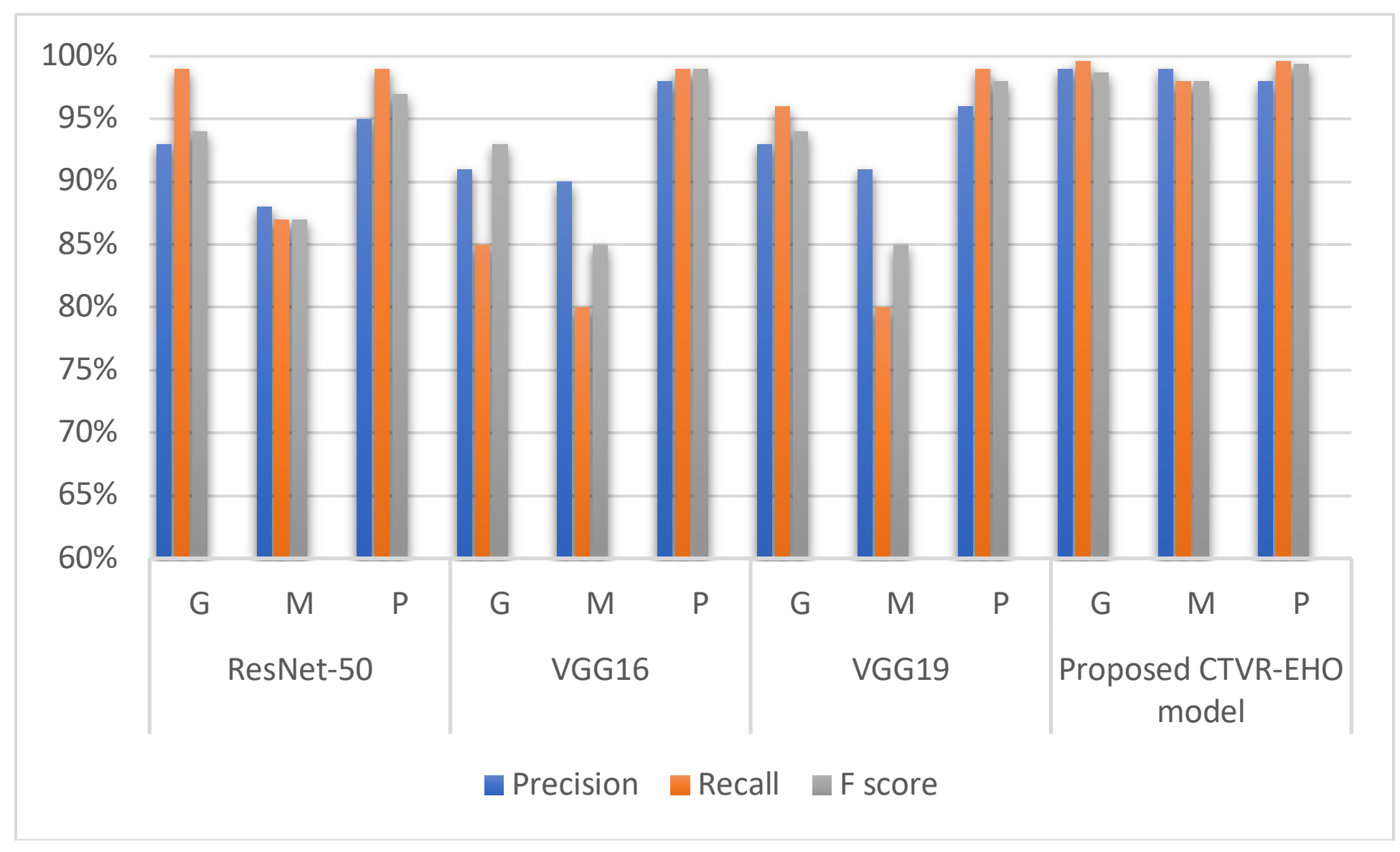

Figure 8: Classification result analysis

Table 10. Comparison of proposed TDA-IPH and CTVR-EHO with existing methods

| Authors | Method | Dataset Used | Precision | Recall | F-score | Accuracy |
|---|---|---|---|---|---|---|
| Maltbie et al. [18] | Restricted sample entropy (RSE) | NHP dataset | 98% | - | - | 98.45% |
| Vidaurre et al. [19] | CV kernel ridge regression | Human Connectome Project | 97.12 | 96.54 | - | 98.78% |
| Ferdous et al. [20] | FCM, GLCM, SVM, KNN | Figshare | 97.02% | 96.5% | 97.8% | 98.11% |
| Shah et al. [21] | Linear-complexity data-efficient image transformer | Figshare and BraTS-21 | - | - | - | 96.11% |
| Ghahfarrokhi and Khodadadi [22] | GLCM and DWT | MRI dataset | - | 89.98% | - | 98.76% |
| Özbay and Özbay [23] | Comparison-to-Learn method, DenseNet201 with Hashing | Figshare, SARTAJ, and Br35H | - | - | - | 98.01% |
| Xu et al. [24] | Cross-visual cortex model and impulse-coupled neural network | Benchmark dataset | - | - | - | 98.99% |
| Deckers et al. [25] | Adaptive filter | MRI dataset | 96.22% | 97.05% | 95.8% | 95% |
| Thillaikkarasi *et al.* [26] | CNN | BRATS 2017 | - | - | - | 84.96% |
| Thaha *et al.* [27] | CNN | BRATS 2019 | 87.78% | 90.02% | 92.92% | 92% |

| Kaldera *et al.* [28] | Faster R-CNN | BRATS 2020 | - | - | - | 94.67% |
|---|---|---|---|---|---|---|
| Sajid *et al.* [29] | Hybrid Convolutional Neural Network | BRATS 2019 | 86.84% | 86.02% | 91.92% | 91% |
| Jia *et al.* [30] | CNN | BRATS 2019 | - | - | - | 98.51% |
| Khan *et al.* [31] | Handcrafted features and CNN | IoT-enabled dataset | 86.94% | 81.92% | 97.12% | 94.66% |
| Mittal *et al.* [32] | Fuzzy C-means + CNN | BRATS 2018 | - | - | - | 98.68% |
| Ozyurt *et al.* [33] | CNN | BRATS 2020 | 96.94% | 94.92% | 93.42% | 98.33% |
| Bhargavi *et al.* [34] | CNN with Element-wise Filters | BRATS 2020 | 67.74% | 66.82% | 68.12% | 69.56% |
| Xing *et al.* [35] | CNN with Element-wise Filters | Brain Networks (custom) | 68.4% | 66.92% | 70.62% | 66.88% |
| Rathore *et al.* [36] | Neural Network based Topological Data Analysis | - | - | - | - | 69.98% |
| Kumar et al. [37] | Residual Network + Global Average Pooling | BRATS 2019 | 97.4% | 97.82% | 96.92% | 97.48% |
| Deepak and Ameer [38] | CNN + SVM | BRATS 2020 | - | - | - | 95.82% |
| Kesav and Jibukumar [39] | RCNN with Two Channel CNN | BRATS 2021 | 97.46% | 95.29% | 98.20% | 98.21% |
| Gupta et al. [40] | MAG-Net | BRATS 2021 | 98% | 98% | 98% | 98.04% |
| Ahuja et al. [41] | DarkNet | Custom dataset | 97.4% | 98.63% | 97.94% | 99.43% |
| Li et al. [42] | LKAU-Net | BRATS 2020 | 96.94% | 96.2% | 96.62% | 98.11% |
| Guan et al. [43] | AGSE-VNet | BRATS 2020 | 95.94% | 97.62% | 98.92% | 97.90% |
| Ghassemi et al. [44] | DNN with GAN Pre-training | BRATS 2020 | 95.94% | 98% | 94.12% | 95.10% |
| Proposed method | TDA-IPH and CTVR-EHO model | Figshare | 99.67% | 99.23% | 99.59% | 99.8% |

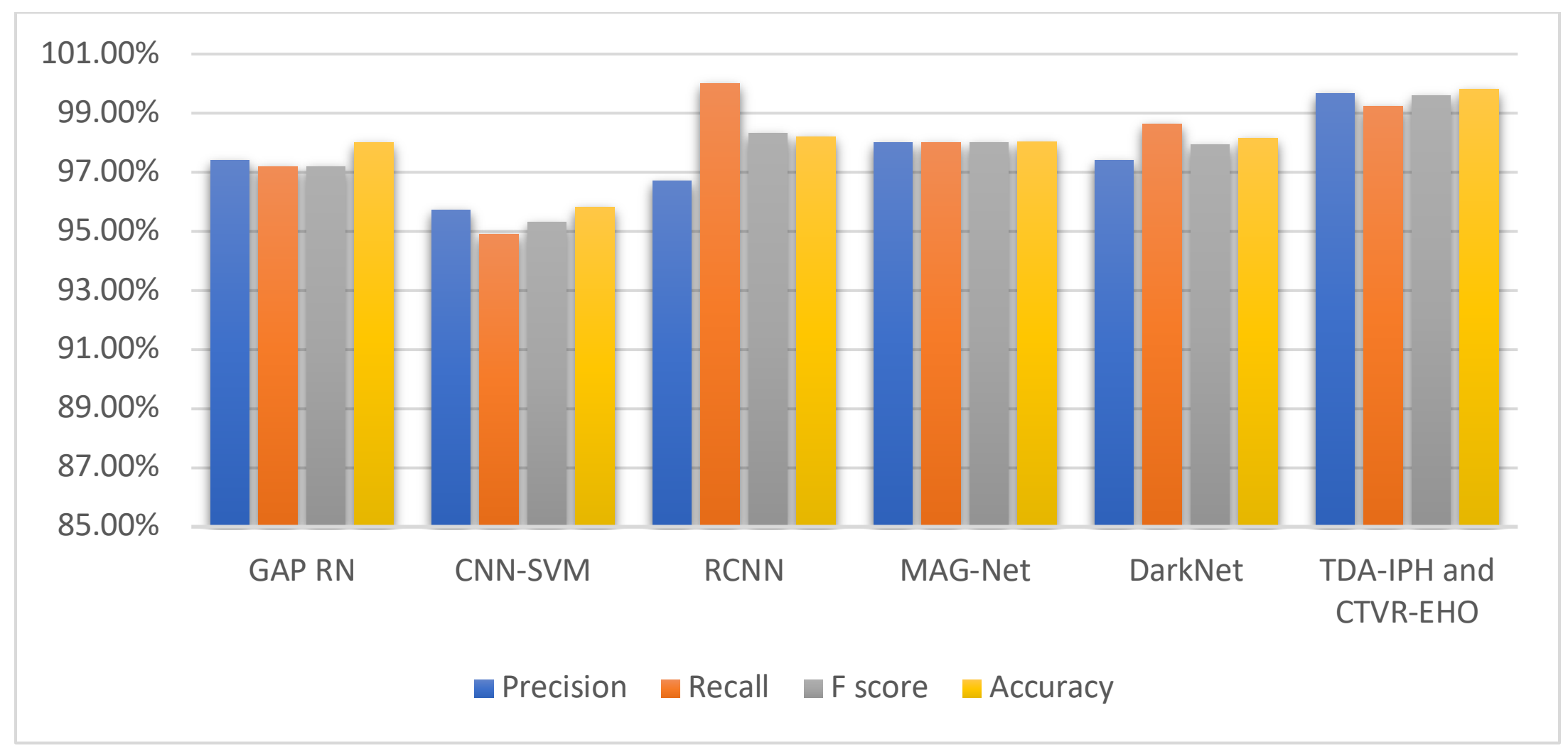

Figure 9: Performance of proposed Comparison of proposed TDA-IPH and CTVR-EHO and existing method

4.5.1 Impact of Training – Validation Accuracy and Loss of Proposed TDA-IPH and CTVR-EHO

The impact analysis of the proposed TDA-IPH and CTVR-EHO model on the Figshare dataset underscores its strong performance in training and validation accuracy and loss, as illustrated in Figure 10 (a). In the early epochs, the model demonstrates rapid learning and effective pattern recognition, quickly capturing significant tumor features. By the 10th epoch, the model's accuracy reaches approximately 99.6%, indicating its adeptness at recognizing inherent patterns within the MRI data. As training progresses, both training and validation accuracy curves converge, the model rapidly learns to distinguish basic tumor characteristics, such as edges and intensity patterns, leading to a sharp increase in accuracy and a decrease in loss. Validation accuracy plateaus at ~94%, demonstrating strong generalization to unseen scans, with a healthy gap between training and validation metrics indicating robust performance without overfitting. In the loss curve shown in Figure 10 (b), the model's training loss exhibits a rapid decline in the initial epochs, starting above 1.2 and reducing sharply to below 0.2 by the 10th epoch. The validation loss follows a similar downward trend, though it stabilizes around 0.2 slightly later. This steady decrease in loss values, coupled with the eventual stabilization, highlights effective learning and robust optimization by the model. Minor fluctuations in the validation loss during later epochs indicate slight variability, yet the overall low and steady values suggest that the model effectively generalizes to new data, efficiently managing the complexities of the Figshare dataset. Therefore, the TDA-IPH and CTVR-EHO model's strong training and validation performance, marked by the convergence of accuracy and reduction in loss, emphasizes its reliability for MRI-based brain tumor segmentation and classification. The stable accuracy and low validation loss ensure the model's suitability for clinical applications, where precision and consistency are paramount for patient care.

Figure 10 (a)& 10(b): Performance results of the TDA-IPH and CTVR-EHO model in terms of training-validation loss and accuracy.

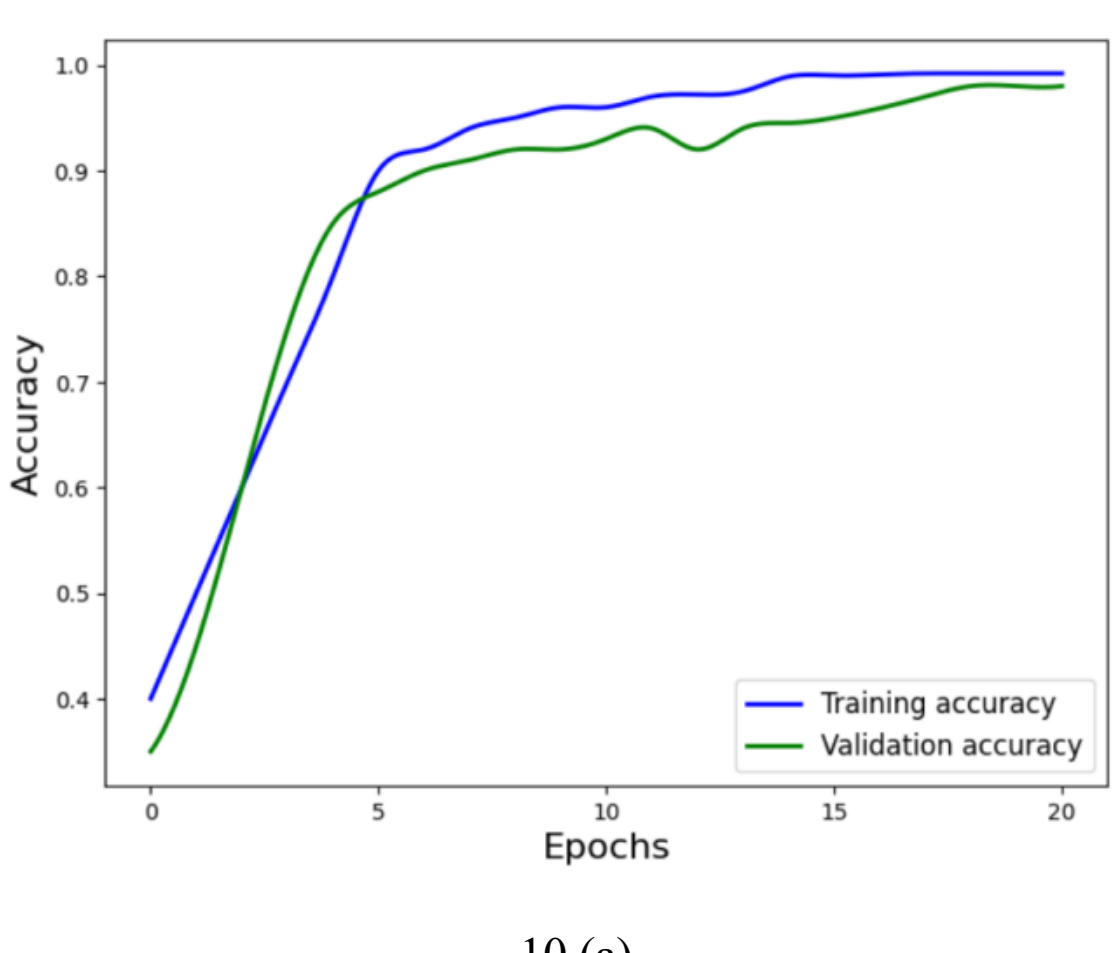

10 (a)

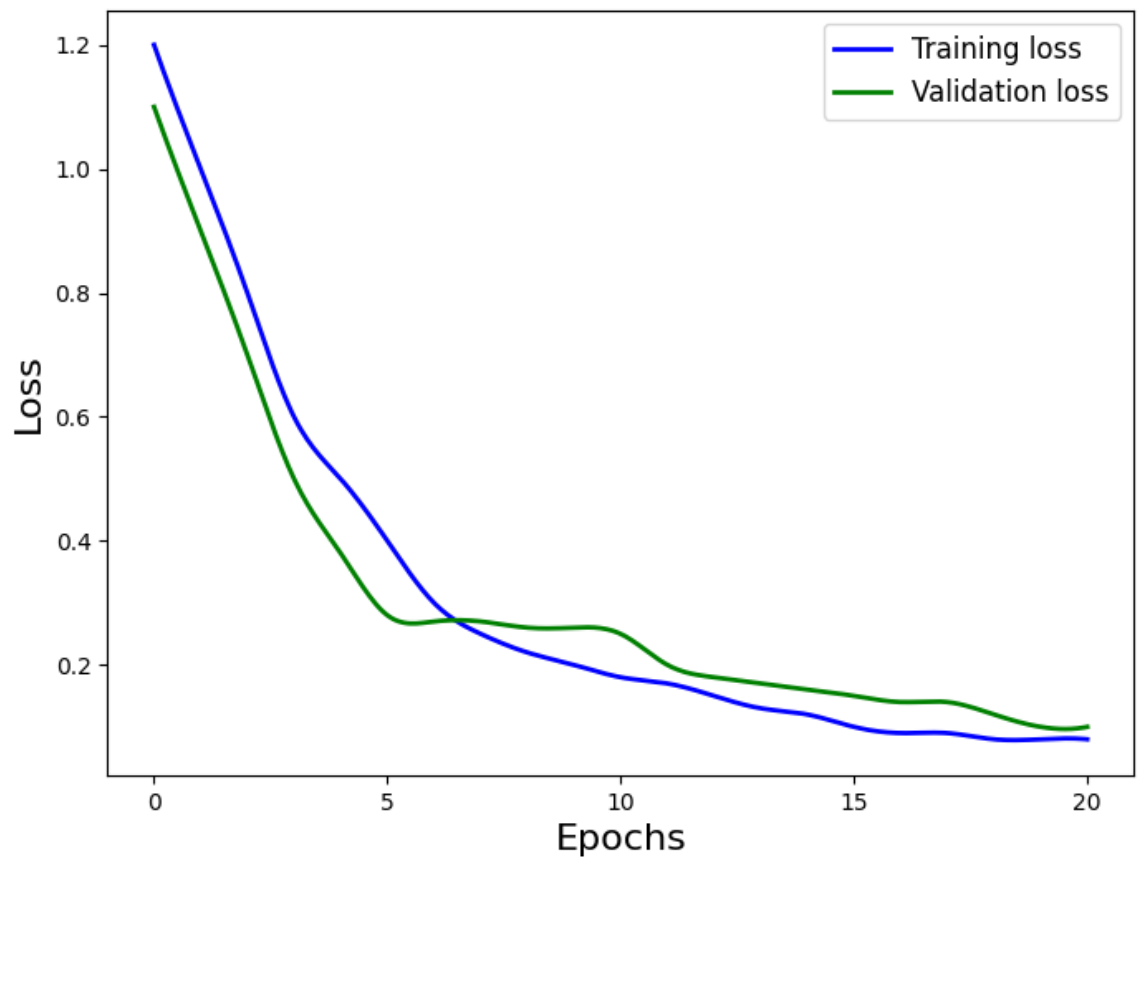

10 (b)

## 5. Conclusion

In this paper, we proposed a novel approach for BT diagnosis using a TDA-IPH based DCTN-BLE model, which can serve as a useful tool for BT diagnosis. The proposed method consists of several steps. Firstly, the MADF algorithm is used to remove noise while preserving the edge from blurring. Then, TDA-IPH is designed to segment the BT image, which separates the porous objects from the image. Next, the CTVR-EHO model is introduced, which extracts the necessary features for BT classification by simultaneously using the AlexNet and BiVLSTM models. The extracted features are concatenated, and BT is classified. Finally, the EHO algorithm is used to tune the hyper-parameters of both AlexNet and BiVLSTM to obtain an optimal result. The performance of our proposed method is evaluated and compared with some prior methods using various metrics. The results show that the proposed TDA-IPH and CTVR-EHO models achieved the best outcome than the existing methods. Our proposed approach can be used to improve the accuracy of BT diagnosis, leading to more effective and efficient treatments.

## 6. Limitations and Future Scope

The proposed method is evaluated using a limited dataset, which may affect the generalizability of our results. Therefore, to ensure the reproducibility of the proposed models for larger scale multi-sequence MRI data or a multi-center study, the data should be collected from multiple centres using standardized imaging protocols and pre-processed in a consistent manner to reduce variability. The data should also be annotated by experienced radiologists to ensure accurate labelling of the tumor regions. We recognize the importance of consulting medical professionals for validation, and while this study did not involve direct consultation with clinicians, we plan to collaborate with radiologists and neurologists in future research to validate our model's findings within a clinical context. This collaboration will enhance the applicability of our approach in real-world diagnostics. Additionally, we aim to explore publicly available datasets that are annotated by experts to further refine our model's accuracy and relevance. The TDA-IPH and CTVR-EHO models should be adapted and optimized for the multi-sequence MRI data and multi-center study. This may involve adjusting the network architecture, hyper-parameters, and training algorithms. Secondly, the proposed method only considered a single modality, i.e., MRI, for BT diagnosis. Combining multiple modalities, such as MRI and PET, may enhance the accuracy of BT diagnosis. Thirdly, while our proposed method achieved promising results, it may still produce false-positive or false-negative results in some cases. Therefore, additional studies are needed to further improve the performance of the proposed method. In future research, we plan to address these limitations and further improve our proposed method. We also aim to extend our approach to other medical imaging applications and explore its potential for clinical translation.

## Ethical Statement

This study does not contain any studies with human or animal subjects performed by any of the authors.

## Conflicts of Interest

The authors declare that they have no conflicts of interest to this work.

## Data Availability Statement

The brain tumor datasets that support the findings of this study are openly available in figshare at https://doi.org/10.6084/m9.figshare.1512427.v5